\documentclass[11pt,preprint,review,12pt,1p,number,sort&compress,a4paper]{elsarticle}

\usepackage[version=3]{mhchem}

\usepackage{mathptmx}
\usepackage{caption}
\usepackage{subcaption}
\usepackage{graphicx}
\usepackage{lastpage}
\usepackage{float}
\usepackage[english]{babel}
\usepackage{array}
\usepackage{droidsans}
\usepackage[usenames,dvipsnames]{xcolor}
\usepackage{hyperref}
\usepackage{chemfig}
\usepackage{tikz}

\usepackage{color}

\usepackage{epstopdf}

\definecolor{cream}{RGB}{222,217,201}

\journal{J. Quant. Spectrosc. Radiat. Transf.}

\begin{document}

\begin{frontmatter}

\title{Collisional broadening of molecular rovibronic lines}

\author{Jeanna Buldyreva$^{a,b}$}
\author{Ryan P. Brady\textit{$^{b}$}}
\author{Sergei N. Yurchenko\textit{$^{b}$}}
\author{Jonathan Tennyson\textit{$^{b}$}}
\address{$^a$~Institut UTINAM UMR CNRS 6213, Universit\'{e} de Franche-Comt\'{e}, 16 Route de Gray, 25030 Besan\c{c}on cedex, France}
\address{$^{b}$~Department of Physics and Astronomy, University College London, Gower Street, WC1E 6BT London, United Kingdom}

\date{\today}

\begin{abstract}
To meet burning needs of high-resolution pressure-induced line-shape parameters in the UV/visible regions for hot-temperature industrial and atmospheric applications as well as current and future space missions,  phase-shift theory is examined in its historical context, tested and revisited using accurate numerical potentials and advanced trajectory models.
First, a general analysis for arbitrary molecular systems is conducted in terms of the dimensionless parameter $\alpha$ determined by the differences of the Lennard-Jones parameters in the final and initial electronic absorber’s states. Temperature dependence, use of the power law and  influence of Maxwell-Boltzmann averaging over relative velocities are addressed. Then, interaction-potential calculations are attempted for some representative molecular pairs (NO-Ar, NO-N$_2$, OH-Ar and OH-N$_2$) and the isotropic parts are fitted using the 12-6 Lennard-Jones form to get room and high-temperature line-broadening and line-shift coefficients which are compared to available measurements. It is shown that the phase-shift theory in its standard rectilinear-trajectory formulation provides linewidth and shift estimates accurate within 30--40\%. Attempted  improvements using numerical
potentials and curved trajectories lead to closer matches with measurements for some cases but also worsen the agreement for others. To ensure better theoretical predictions, introduction of correction terms to the usual phase-shift integral is suggested.

\end{abstract}







\end{frontmatter}

\newpage

\section{Introduction}

A spectroscopic transition is specified by three elements: a line centre, a line intensity and a line shape. Much effort has been expended
characterising line centres and intensities but the line shape can also be crucial for determining overall absorption,
for example by the key CO$_2$ molecule in the terrestrial atmosphere \cite{22RoSeJa.CO2}. While the theory of molecular line shapes
for rotation-vibration transitions which generally lie in the infrared is relatively well developed \cite{08HaBoRo,10BuLaSt,18HaTrAr}, significantly less attention has been paid to rovibronic molecular transitions which are usually observed at visible or
ultraviolet wavelengths. Electronic transitions of molecules such as OH, NO and O$_3$ are important in the Earth's
atmosphere while a variety of such transitions are crucial in technological applications \cite{23Fateev}, including
combustion \cite{Westlye2022}. Of particular interest to us is the importance for astronomy, namely for applications to studies
of objects such as brown dwarfs and exoplanets. The standard transit method of obtaining exoplanet spectra, as currently being employed, for example, by both the Hubble Space Telescope and the James Webb Space Telescope, by construction probes line spectra
to the point of optical thickness in the atmospheres of exoplanets. For optically thick lines it is the line profile
that largely determines the overall opacity of a line. Test models of exoplanets have shown their sensitivity to the inclusion of
line broadening effects \cite{jt512,jt866} while at the same time there is a general awareness of the lack of line broadening data for such studies \cite{19FoRoDo}. Transport of visible and ultraviolet radiation in exoplanet atmospheres is a subject of increasing study and it therefore seems appropriate to investigate closely the molecular line broadening effects at these wavelengths.


It is now a well-established fact that the pressure-broadened linewidths of vibrotational and pure rotational transitions observed in the infrared (IR) and microwave (MW) frequency regions are mainly influenced by inelastic collision processes \cite{DeLucia1988} and exhibit (except for very light collision partners such as helium and hydrogen) a pronounced dependence on the rotational quantum number $J$. Theoretical interpretation of such linewidths and associated shifts includes adiabatic (molecular axis reorientation and vibrational dephasing) and non-adiabatic (state change) contributions.

The situation is completely different if the spectroscopically active molecule jumps from its ground electronic state to an excited electronic state. Such transitions form bands of rovibronic lines which lie generally at the ultraviolet (UV) and/or visible wavelengths; however, we note that some important astronomical species such as TiO, VO, CrH, FeH and CN have electronic
bands which absorb strongly in the near infrared \cite{jt731}. In contrast to pure vibrational bands, the pressure-broadened linewidths of vibronic bands lack observable $J$-dependence which argues in favour of the minor importance of rotational energy transfer.

In this paper we explore use of phase-shift theory for providing pressure-dependent line-shape parameters for rovibronic spectra.
This simple classical approach is chosen because of intended high-temperature applications, in particular to exoplanetary atmospheres where many ``exotic'' molecular pairs are detected or expected but even approximate and rotationally-independent collisional line-shape parameters are completely missing.
The next section describes a historical context of this theory to get clearer for the general reader fundamental differences and working formulae of theoretical treatments of rovibronic and rovibrational pressure line-broadenings. Section 3 gives the basic methodology and the results of its testing with traditionally modelled and numerical intermolecular interaction potentials.
Section 4 focuses on possible improvements of the trajectory model and Section 5 gives our conclusions.

\section{Main steps in theoretical developments}
Theoretical descriptions of the rovibronic line-shape parameters take their roots in primary approaches developed at the beginning of the twentieth century.
The earliest treatment of spectral linewidths induced by collisions of a radiating atom/molecule with neighbouring particles is due to Lorentz \cite{Lorentz1906} who used a classical radiating oscillator model with collisions changing the phase by an arbitrary amount and resulting, via  Fourier integral analysis, in the ``dispersion'' (Lorentzian) line shape. The half-width at half-maximum (HWHM)  was expressed as a function of the mean relative speed of the colliding molecules, the number of molecules in a unit volume and the molecular ``radius'' considered as a mere parameter (its value was found to be very different from the kinetic collision radius). Weisskopf \cite{Weisskopf1932} pushed further Lorentz’ idea of the classical oscillator assuming that its frequency varies in time during the collision process. However, his assumption that ``big-magnitude'' ($>1$) phase shifts are equivalent to arbitrary phase shifts and that smaller phase shifts are negligible excluded consideration of line-centre shifts.

Another (static) approach was attempted later by Kuhn and London \cite{KuhnLondon1934} and by Margenau \cite{Margenau1935, Margenau1936} who considered the active particles as randomly distributed in fixed positions and simultaneously interacting with all the perturbing atoms. The line shape obtained by Kuhn and London gave a diverging intensity at the line centre and that obtained by Margenau resulted in the widthand shift proportional to the squared gas density. A more rigorous approach was given by Jablonski \cite{Jablonski1937, Jablonski1940, Jablonski1945} who considered the gas as analogous to a very large molecule and calculated the stationary states of internal motion.
Making an analogy with the  Franck-Condon principle \cite{Foley1946} and averaging the expression for a single perturber over all possible transitions, he obtained a formula equivalent to that of Kuhn and London. The alternative approaches explained the asymmetry of line shapes observed at high pressures but were inappropriate for low pressures.

Jablonski attributed the disagreement of his expression with the Lorentz-Weisskopf line shape to the non-validity of the Fourier-integral method, but Lindholm \cite{Lindholm1945} and Foley \cite{Foley1946} successfully developed theories of this type. Namely, Foley showed that the quantum radiation theory with an adiabatic collision approximation and the actual distribution of phase shifts yields a Fourier-integral expression for the intensity distribution with both linewidth and shift linearly proportional to the pressure. He also argued that non-adiabatic effects are small for radiating atoms because of large separations between their levels, but they can occur for molecular IR spectra due to close energy gaps between the rotational levels of the active and perturbing molecules.

The quantum-mechanical adiabatic approach initiated by Jablonski \cite{Jablonski1937, Jablonski1945} was further developed by Baranger \cite{Baranger1958} and other authors \cite{Royer1972a,Royer1972b, Fiutak1970a,Fiutak1970b, Sando1973, Sando1974, Futrelle1972}. A unified model based on the Franck-Condon principle and yielding both impact (valid for the line core) and quasistatic (applying in the wings) approximations in the appropriate limits was suggested by Szudy and Baylis \cite{Szudy1975}. These authors assumed that  the perturbing particles move independently (the binary collision regime) in adiabatic potentials $V_i (r)$ and $V_f(r)$ corresponding to the initial and final electronic states of the radiating atom and pointed out that in classical treatments there is uncertainty in the choice of the potential energy surface (whether it should correspond to the initial state, to the final state or to some mixture) and that this uncertainty is frequently avoided using straight-line trajectories independent of the adiabatic potential. Their results indicated that for a ``reasonable path'' the radial velocity at any trajectory point should be ``the average of the initial- and final-state radial velocities'' with ``abrupt reversal of radial velocity''  at the turning point. It means that the usual phase-shift integral should be improved by, at least, a first-order correction term, and significant differences can be obtained with respect to calculations with straight-line trajectories. In the classical limit the authors recovered the Lindholm-Foley \cite{Lindholm1945, Foley1946} expressions for the linewidth and shift, and calculated the first-order correction term for an inverse-power potential with straight trajectories to get the unified line-shape with an asymmetry effect.

Some authors (see, e.g., refs. \cite{Mizushima1967, Berman1972, Ward1974} for atomic radiators and \cite{Luijendijk77, Pickett1980} for absorbing/emitting molecules) also addressed the velocity-dependence of collisional broadening, closely related to the temperature dependence of collisional linewidths, and analysed the procedure for averaging over relative molecular speeds which leads to departures from the mono-velocity Lorentz shape and the appearance of an additional linewidth parameter characterizing its speed dependence.

Theories of collisional broadening of Lorentz-shape molecular spectral lines have been developing since the late 1940s in two different ways. One of them concerned a description of MW/IR lines corresponding to rotational/vibrotational transitions for which the major effect of inelastic collisions had been evidenced by Anderson \cite{Anderson1949} and Tsao and Curnutte \cite{Tsao1962} (so called Anderson-Tsao-Curnutte theory). In this semi-classical approach, based on perturbation theory and the impact approximation, two colliding molecules separated by the time-dependent distance $r$ follow straight-line trajectories in their relative motion and interact via long-range forces. As the active molecule remains in its ground electronic state, the same intermolecular potential surface is used for the initial and final states of the observed transition. Linewidths predicted from a (usually modelled) interaction potential have much higher values than shifts, in full agreement with measurements. Non-perturbative treatments employing classical trajectories were suggested also by Neilsen and Gordon \cite{Neilsen1973a,Neilsen1973b} and by Smith et al. \cite{Smith1976} A short summary of evolution of different theories up to 1980 was published by Leavitt \cite{Leavitt1980}, and updated reviews (including fully quantum approaches) were given by Boulet \cite{Boulet2004} and by Hartmann et al. \cite{18HaTrAr}.

Other authors such as Mizushima \cite{Mizushima1951} continued developing the phase-shift approach (intended at that time also for pure rotational transitions) with the same assumption of the second-order perturbation, straight-line trajectories and one leading term of $r^{-n}$ type in the interaction potential. The practically important cases of $n$ = 3, 5 and 6 corresponding to dipole-dipole, quadrupole-quadrupole and dispersion interactions, respectively, were considered for both widths and shifts (having the same order of magnitude), and, under assumption of identical shift signs for all collisions, Foley’s relation for the ratio of absolute shift to width values equal to $\tan[\pi/(n-1)]$ was retrieved. The expressions for collisional linewidth and shift derived by Mizushima contained explicit temperature dependence and served as a basis for the widely used one-power law for linewidths (a similar law for shifts has been finally invalidated by measurements). Although Mizushima obtained qualitative agreement with measured linewidths for some linear molecules, further studies of microwave and infrared absorption unambiguously demonstrated the inappropriateness of phase-shift approaches to collisional broadening and shifting of spectral lines associated with rotational and vibrotational transitions. An evaluation of phase-shift theories was given by Breene \cite{Breene1961}.

It was underlined by Margenau \cite{Margenau1936-PhysRev} that lines of an electronic transition, i.e. lines of UV/visible absorption bands, are broadened and shifted similarly to atomic lines corresponding to the type and electronic energy jump. Therefore, classical phase-shift theories offer an alternative to a quantum-mechanical formulation. The hypothesis of adiabatic collisions, valid for radiating atoms,  appears to be  well justified for rovibronic lines of the A\ $^2\Sigma^+\leftarrow$~X\ $^2\Pi$ (0,0) band (the so-called $\gamma$ band) of NO perturbed by Ar and N$_2$ at 295 and 2800~K \cite{Chang1992} as well as for the B\ $^1\Sigma^+ \leftarrow$ X\ $^1\Sigma^+$ (0,0) band of CO perturbed by N$_2$, CO$_2$, CO at 294, 656 and 1010~K \cite{DiRosa2001} which show no $J$-dependence for either widths or shifts. Also, as pointed out by Di Rosa and coworkers \cite{DiRosa1994}, generally very different the upper and lower electronic states mean that the broadening and shifting of spectral lines are likely induced only by dispersive forces for neutral and non-polar (weakly polar) perturbers; the measured red (negative) shifts \cite{DiRosa1994} argue in favour of the dominance of dispersion forces too. In general, the magnitudes of linewidths and shifts of rovibronic transitions are bigger than those of (vib)rotational transitions and cannot be explained by IR/MW line-broadening theories assuming the major contribution from inelastic collisions.

However, it should be noted that the application of phase-shift theories needs care because of important simplifications made: only isotropic interactions cannot account for orientational dependence of elastic transitions (no $J$-dependence can be predicted) and inelastic interactions are excluded. Birnbaum \cite{Birnbaum1967-InterForces} and Thorne \cite{Thorne1974} pointed out that inelastic transitions can contribute to the broadening but not to the shift of rovibronic transitions through either electrostatic interactions (which include resonance interactions) or inductive interactions or both. These interactions induce non-radiative transitions and for some active molecules can create a $J$-dependence: e.g., for the A\ $^2\Sigma^+ \leftarrow$ X\ $^2 \Pi(0,0)$ band of OH perturbed by Ar and N$_2$ in the temperature range 1400--4100~K \cite{Rea1987} the measured linewidths (1.5$\leq J\leq$17.5) exhibited nearly negligible $J$-dependence, whereas the R$_1$-branch lines of the same band influenced by H$_2$O and CO$_2$ pressure over the range 1470--2370~K \cite{Rea1989} demonstrated some slight dependence (for H$_2$O-perturbation the widths are also ``resonantly enhanced'' because of the strong dipole moment of this perturber).

Table~\ref{Table1} gives an idea of the magnitude and temperature-dependence of the parameters measured for molecular systems: linewidth and shift data as well as associated temperature exponents (extracted using the standard power law) reported in the literature for rovibronic transitions of NO, CO and OH perturbed by various gases. The most extensive data set is for  NO due to its chemical stability at room and elevated temperatures. For the NO case, the data refer to the $\gamma$(0,0)-band but they can be also used for the $\gamma$(1,0)-band up to 1700~K \cite{Trad2005}. Moreover, Di Rosa et al. \cite{DiRosa1996} noted that the widths and the temperature exponents for NO-N$_2$ work well for NO-CO and NO-CO$_2$, so that N$_2$-broadening parameters are reproduced for CO and CO$_2$ perturbers. The use of temperature-dependence parameters for NO-N$_2$ furnished successful simultaneous measurements of velocity, pressure and temperature in planar laser-induced fluorescence of nitric oxide \cite{Naik2009}. Very recent measurements by Krish et al. \cite{Krish2022} for NO-Ar mixtures in the high-temperature range 2000--6000~K suggest the use of two sets of linewidth and temperature-exponent parameters for 2000--2500~K and 2500--6000~K (with the reference temperature of 296~K). According to the theoretical analysis by Cybulski and co-authors \cite{Cybulski2013},  phase-shift theory accounting for both long-range (e.g., dispersive) and short-range (repulsive wall) explains such influence of the temperature range by the leading contributions from different (attractive or repulsive) forces. The absence of any $J$-dependence for the widths and shifts of NO lines perturbed by N$_2$ and Ar (small oscillator strength makes resonance effects improbable and the dipole moment of NO is small) but also by H$_2$O and O$_2$ (with polarizabilities close to that of Ar and N$_2$) makes NO a molecule suitable for the phase-shift theory. For CO, quite extensive measurements showing no dependence on rotational states were published by Di Rosa et al. \cite{DiRosa2001} for self-perturbation and perturbation by N$_2$ and CO$_2$ at temperatures in the interval 294--1010~K. For OH, the measurements performed for Ar- and N$_2$-broadenings \cite{Rea1987} showed no pronounced $J$-dependence, with N$_2$ inducing slightly larger linewidths; the $J$-averaged data \cite{Rea1987} are reproduced in Table~\ref{Table1}. Absence of observed $J$-dependence was also reported earlier by Engleman \cite{Engleman1969} who studied perturbation by He, Ne, Ar, Kr, Xe, H$_2$, D$_2$, O$_2$, N$_2$, CH$_4$, CO$_2$, N$_2$O, SF$_6$, CF$_4$ at 293~K and by H$_2$O at 378~K and concluded that linewidths are independent of $J$ for all gases except water and N$_2$O, for which cross-section was found to be a decreasing function of $J$. A subsequent study by Hwang et al. \cite{Hwang2008} of the P$_1$(5) line of A\ $^2\Sigma^+ \leftarrow$ X\ $^2\Pi$ (0,0) band of OH perturbed by Ar, N$_2$ and H$_2$O (780--2440~K, 0.7--10.0~atm) revealed that at high temperatures short-range repulsive interactions can play a role and the need for quantum-mechanical calculations was evoked, since the current state of theoretical treatment is not adequate.

\begin{table}
\scriptsize
  \caption{\ Collisional broadening and shift coefficients (in cm$^{-1}$atm$^{-1}$) and temperature exponents for linewidths ($n$) and shifts ($q$) measured for vibronic transitions.}
  \label{Table1}
  \begin{tabular*}{\textwidth}{@{\extracolsep{\fill}}llllllllll}
    \hline
   & HWHM & shift & $T$~(K) & $T_{\rm ref}$ (K) & HWHM $T_{\rm ref}$  & $n$  & shift $T_{\rm ref}$ & $q$ & ref. \\
\hline
NO-N$_2$ &  0.348 & - & 295 & - & - & - & - & - & \cite{Dodge1980} \\
         &  0.292 & -0.180(5) & 295--2700 & 295 & 0.293 & 0.75(5) & -0.180(5) & 0.56 & \cite{Chang1992} \\
         &  0.293(20) & - & ``room'' & - & - & - & - & - & \cite{Danehy1993} \\
         &  0.279 & - & 300 & - & - & - & - & - & \cite{Hanna2002} \\
         &  0.293 & - & 300 & - & - & - & - & - & \cite{Anderson2005} \\
         &  0.291(3) & -0.174(2) & 296 & - & - & - & - & - & \cite{Shao2007} \\
\hline
NO-Ar &  0.269 & - & 295 & - & - & - & - & - & \cite{Dodge1980} \\
      &  0.252 & -0.159 & 295--2800 & 295 & 0.253 & 0.65(3) & -0.160 & 0.58(3) & \cite{Chang1992} \\
      &  0.25 & -0.16 & 295 & - & - & - & - & - & \cite{DiRosa1994} \\
      &  -    & -     & 2000--2500 & 296 & 0.175 & 0.65 & - & - & \cite{Krish2022} \\
      &  -    & -     & 2500--6000 & 296 & 0.370 & 1 & - & - & \cite{Krish2022} \\
\hline
NO-CO$_2$&  -    & -     & - & 295 & 0.293 & 0.75 & - & - & \cite{DiRosa1996} \\
         &  0.303(10) & - & "room" & - & - & - & - & - & \cite{Danehy1993} \\
\hline
NO-CO &  -    & -     & - & 295 & 0.293 & 0.75 & - & - & \cite{DiRosa1996} \\
\hline
NO-O$_2$&  0.265 & -0.16 & 295 & 295 & - & 0.66*** & - & - & \cite{DiRosa1994} \\
        &        &       &     & 273.2 & 0.105 & 0.7 & -0.063 & 0.7 & \cite{DiRosa1996} \\
\hline
NO-H$_2$O&  0.395 & -0.21 & 295 & 295 & - & 0.79*** & - & - & \cite{DiRosa1994} \\
\hline
NO-NO &  0.276 & -0.171 & 295 & - & - & - & - & - & \cite{Chang1992} \\
      &  0.275 & -0.17 & 295 & - & - & - & - & - & \cite{DiRosa1994} \\
\hline
CO-N$_2$ &  0.36(1) & -0.215(2) & 294(1) & 295 & 0.365(15) & 0.77(7) & -0.22(1) & 0.53(4) & \cite{DiRosa2001} \\
         &  0.215(5) & -0.150(4) & 656(15) & - & - & - & - & - &  \\
         &  0.140(5) & -0.109(4) & 1010(20) & - & - & - & - & - &  \\
 \hline
CO-CO$_2$ &  0.39(1) & -0.166(9) & 294(1) & 295 & 0.385(15) & 0.63(7) & -0.17(1) & 0.75(9) & \cite{DiRosa2001} \\
          &  0.23(1) & -0.102(7) & 656(15) & - & - & - & - & - &  \\
          &  0.180(5) & -0.065(8)& 1010(20) & - & - & - & - & - &  \\
 \hline
CO-CO &  0.37(1) & -0.211(3) & 294(1) & 295 & 0.37(2) & 0.65(8) & -0.21(1) & 0.52(1) & \cite{DiRosa2001} \\
      &  0.230(5) & -0.138(3) & 656(15) & - & - & - & - & - &  \\
      &  0.160(5) & -0.111(1)& 1010(20) & - & - & - & - & - &  \\
\hline
OH-N$_2$ &  - & - & 1500--2500 & 2000 & 0.02* & 0.67 & - & - & \cite{Rea1987} \\
         &  - & - & 783--2434 & 1000 & - & 0.75 & - & - & \cite{Hwang2008} \\
 \hline
OH-Ar &  - & - & 1400--4100 & 2000 & 0.018* & 0.94 & - & - & \cite{Rea1987} \\
         &  - & - & 1268--2441 & 1000 & - & 1.0 & - & - & \cite{Hwang2008} \\
 \hline
OH-H$_2$O&  - & - & 1620, 2370 & 1620 & - & [-0.1,1] & - & - & \cite{Rea1989} \\
         &  - & - & 783--2434 & 1000 & - & 0.87 & - & - & \cite{Hwang2008} \\
 \hline
OH-CO$_2$&  [0.024,0.054] & - & 1680 & ** & ** & [1.2,2.2] & - &  - & \cite{Rea1989} \\
         & [0.012,0.038]    & - & 2290 & **& ** & - & - & - &  \\
    \hline
  \end{tabular*}
  * Mean value estimated from Figs 4 and 5 of \cite{Rea1987}\\
  ** Not communicated by the authors \cite{Rea1987} because of too large values obtained for $n$ and important overall scatter between successive $J$-values\\
  *** Values reported in \cite{Trad2005}.
\end{table}

Whereas the validity of the simple power law, at least in limited temperature intervals, is well established for linewidths, its application to shifts is questionable because of the sign change observed when going from low to high temperatures. This effect was observed for IR transitions (see, e.g., Baldacchini et al. \cite{Baldacchini1996} who reported shifts changing from red at 200~K to blue at 285 and 315~K for 2 studied lines of NH$_3$) and MW transitions (see Ulivi et al. \cite{Ulivi1989} who observed changes from red shifts at 77~K to blue at 195~K and 296~K for HD). For radiating atoms, a similar change of shift sign (red for $T<$~500~K, small blue for $T>$~500~K) was observed by Bobkowski et al. \cite{Bobkowski1986, Bobkowski1987} for Ne; this change was explained by Findeisen et al. \cite{Findeisen1987} within the framework of the phase-shift theory as appearing at higher temperature if the repulsive interaction is accounted for. It means that a correct theoretical description should include both long- and short-range interactions to properly mimic line-broadening and line-shifting mechanisms from low to high temperatures. We note also that the phase-shift-theory frame is approximate, and improvements of the one-power law are out of interest and scope of the present study.

Given that the line-shape parameters have been generally observed to be independent of $J$ for a large number of molecular systems, below we revisit basic phase-shift theory as developed by Mizushima \cite{Mizushima1951} (who accounted for velocity averaging in his leading-interaction-term approach) and Hindmarsh et al. \cite{Hindmarsh1967} (who used the mean thermal velocity but took account of both attraction and repulsive forces in the form of 12-6 Lennard-Jones potential model).
Like these authors, we start with the common model of straight-line trajectories but consider both the mean-thermal-velocity approximation (MTVA) and Maxwell-Boltzmann averaging (MBA) on relative velocities (as did Cybulski et al. \cite{Cybulski2013} in their general theoretical analysis for Lennard-Jones 12-$m$ interactions). Then, we develop models of curved trajectories governed by the isotropic interactions in the initial and final states of the optically active molecule and discuss the influence of this modified trajectory treatment on the calculated broadening/shift coefficients and their temperature dependence for the test systems NO-Ar and NO-N$_2$.
Even with no $J$-dependence considered, we keep the term ``rovibronic'' for the sake of coherence with papers published on this subject and for underlying the fact that an electronic-state change is accompanied by a (vibration-)rotation-state change.

\section{Phase-shift theory with straight-line trajectories: general analysis}
\subsection{Mean-thermal-velocity approximation}

Within the MTVA, the phase-shift-theory formulae for collisional linewidth $\gamma$ and shift $\delta$ (in s\textsuperscript{-1}) are
\begin{equation}
\label{eq1}
\gamma = N\overline{v}\int_{0}^{\infty}{\left\lbrack 1 - \cos{\eta(b)} \right\rbrack bdb}\ ,
\end{equation}
\begin{equation}
\label{eq2}
\delta = N\overline{v}\int_{0}^{\infty}{\sin{\eta(b)}bdb}
\end{equation}
with \emph{N} denoting the number of molecules per unit volume,
\(\overline{v}\) standing for the mean thermal velocity and \(\eta(b)\)
being the phase shift induced in the radiation by a collision of impact
parameter \emph{b}. The phase shift \(\eta(b)\) represents an
accumulation of the frequency displacements \(\Delta\omega\) (in
rad~s\textsuperscript{-1}) at time \emph{t} through the trajectory:
\begin{equation}
\label{eq3}
\eta(b) = \int_{- \infty}^{+ \infty}{\Delta\omega dt}
\end{equation}
and $\Delta \omega$ itself is determined by the intermolecular interactions inversely proportional to the $n$-th powers of the
intermolecular distance $r$.

Whereas only leading terms were considered by many authors thus resulting in analytic final expressions for $\gamma$ and $\delta$ (see, e.g.,  \cite{Mizushima1951}), the theoretical approach developed by Hindmarsh and coauthors \cite{Hindmarsh1967} suggested a much more realistic treatment via a combination of both attractive and repulsive forces in a Lennard-Jones 12-6 form:
\begin{equation}
\label{eq4}
\eta(b) = \int_{- \infty}^{+ \infty}\left\lbrack \frac{\Delta C_{12}^{\prime}}{r^{12}(t)} - \frac{\Delta C_{6}^{\prime}}{r^{6}(t)} \right\rbrack dt \ ,
\end{equation}
where the parameters $\Delta C_{12}$ =$ \hbar \Delta C_{12}^{\prime}$ and
$\Delta C_{6} = \hbar \Delta C_{6}^{\prime}$ refer to the
\emph{differences} between the Lennard-Jones intermolecular potential
parameters for the final and initial states of the transition considered
(so that both positive and negative signs are possible for \(\Delta C\)
values).

\subsubsection{Trajectory integrals}

The use of straight-line trajectories with
\[ r^{2}(t) = b^{2} + {(\overline{v}t)}^{2} \] for the relative molecular
motion readily transforms Eq.~(\ref{eq4}) into
\begin{equation}
\label{eq5}
\eta(b) = \alpha x^{- 11} - x^{- 5} \ ,
\end{equation}
where the short-hand notations
$$ x = b\left( \frac{8\overline{v}}{3\pi\left| \Delta C_{6}^{\prime} \right|} \right)^{1/5}
$$
and
$$\alpha = \frac{7{\overline{v}}^{6/5}\Delta C_{12}^{'}}{2^{7/5}3^{1/5}\pi^{6/5}\left| \Delta C_{6}^{\prime} \right|^{11/5}}
$$
are introduced. The half-width and shift can therefore be written as
\begin{equation}
\label{eq6}
\gamma = 2\left( \frac{3\pi}{8} \right)^{2/5}\left| \Delta C_{6}^{\prime} \right|^{2/5}{\overline{v}}^{3/5} N B(\alpha) \ ,
\end{equation}
\begin{equation}
\label{eq7}
\delta = \left( \frac{3\pi}{8} \right)^{2/5}\left| \Delta C_{6}^{\prime} \right|^{2/5}{\overline{v}}^{3/5} N S(\alpha)
\end{equation}
depending on the so-called trajectory integrals
\begin{equation}
\label{eq8}
B(\alpha) = \int_{0}^{\infty}\sin^{2}\left[\frac{1}{2}(\alpha x^{- 11} - x^{- 5})\right]xdx \ ,
\end{equation}
\begin{equation}
\label{eq9}
S(\alpha) = \int_{0}^{\infty}\sin(\alpha x^{- 11} - x^{- 5})xdx \ .
\end{equation}
By the change of variables $ y \equiv x^{-1}$ the integrals in Eqs.~(\ref{eq8}) and (\ref{eq9}) take the forms
\begin{equation}
\label{eq10}
B(\alpha) = \int_{0}^{\infty}\sin^{2}\left[ \frac{1}{2}\left( \alpha y^{11} - y^{5} \right) \right]y^{-3} dy \ ,
\end{equation}
\begin{equation}
\label{eq11}
S(\alpha) = \int_{0}^{\infty}\sin(\alpha y^{11} - y^{5})y^{- 3}dy
\end{equation}
and can be expressed analytically but their very lengthy expressions contain special functions.
On the other hand, direct numerical integration of Eqs.~(\ref{eq10}) and (\ref{eq11}) is time-consuming and unstable (incorrect limits have been found for $\alpha \rightarrow 0$). One integration by parts can be performed for $B(\alpha)$ and $S(\alpha)$ of Eqs.~(\ref{eq8}) and (\ref{eq9}) to get more tractable integrals for numerical evaluation:
\begin{equation}
\label{eq12}
B(\alpha) = \frac{1}{4}\int_{0}^{\infty}{\sin\left( \alpha x^{- 11} - x^{- 5} \right)(11\alpha x^{- 10} - 5x^{- 4})} dx \ ,
\end{equation}
\begin{equation}
\label{eq13}
S(\alpha) = \frac{1}{2}\int_{0}^{\infty}{\cos\left( \alpha x^{- 11} - x^{- 5} \right)(11\alpha x^{- 10} - 5x^{- 4})} dx \ .
\end{equation}
The case of a 12-6 Lennard-Jones interaction corresponds to a leading $m=6$ term of the dispersive forces acting between two particles and contributing to the isotropic part of the intermolecular potential. The early work of Mizushima \cite{Mizushima1951} (dealing with one-term interactions) considered also interactions varying as $m=5$ and $m=3$ inverse powers of the intermolecular distance, as he aimed at the description of microwave spectra due to dipole-dipole and quadrupole-quadrupole interactions. Cybulski and co-authors \cite{Cybulski2013} also addressed the cases $m=5$ and $m=4$ with the goal of checking the applicability of the one-power law for the temperature dependence of broadening and shifting coefficients for quadrupole-quadrupole and dipole-quadrupole interactions. However, it should be kept in mind that the isotropic interactions determining the linewidth and shift in the framework of phase-shift theory, besides the repulsive wall, come from the dispersive part of long-range forces which are represented by  the sum of   $r^{-6}$, $r^{-8}$, $r^{-10}$, etc. contributions. Therefore, we do not have to deal with  $ m < 6$ for our goal of estimating broadening and shift parameters for vibronic transitions. Nevertheless, for the sake of comparisons with the previously published results \cite{Mizushima1951, Cybulski2013} we provide the bulk of definitions of $\alpha$, $x$, $B(\alpha)$, $S(\alpha)$ for Lennard-Jones 12-$m$ interactions ($m$=~6, 5, 4) and the corresponding expressions for linewidths and shifts (Tables~\ref{TableA1} and \ref{TableA2} in Appendix).

\begin{figure}[!ht]
    \centering
    \includegraphics[width=0.55\linewidth]{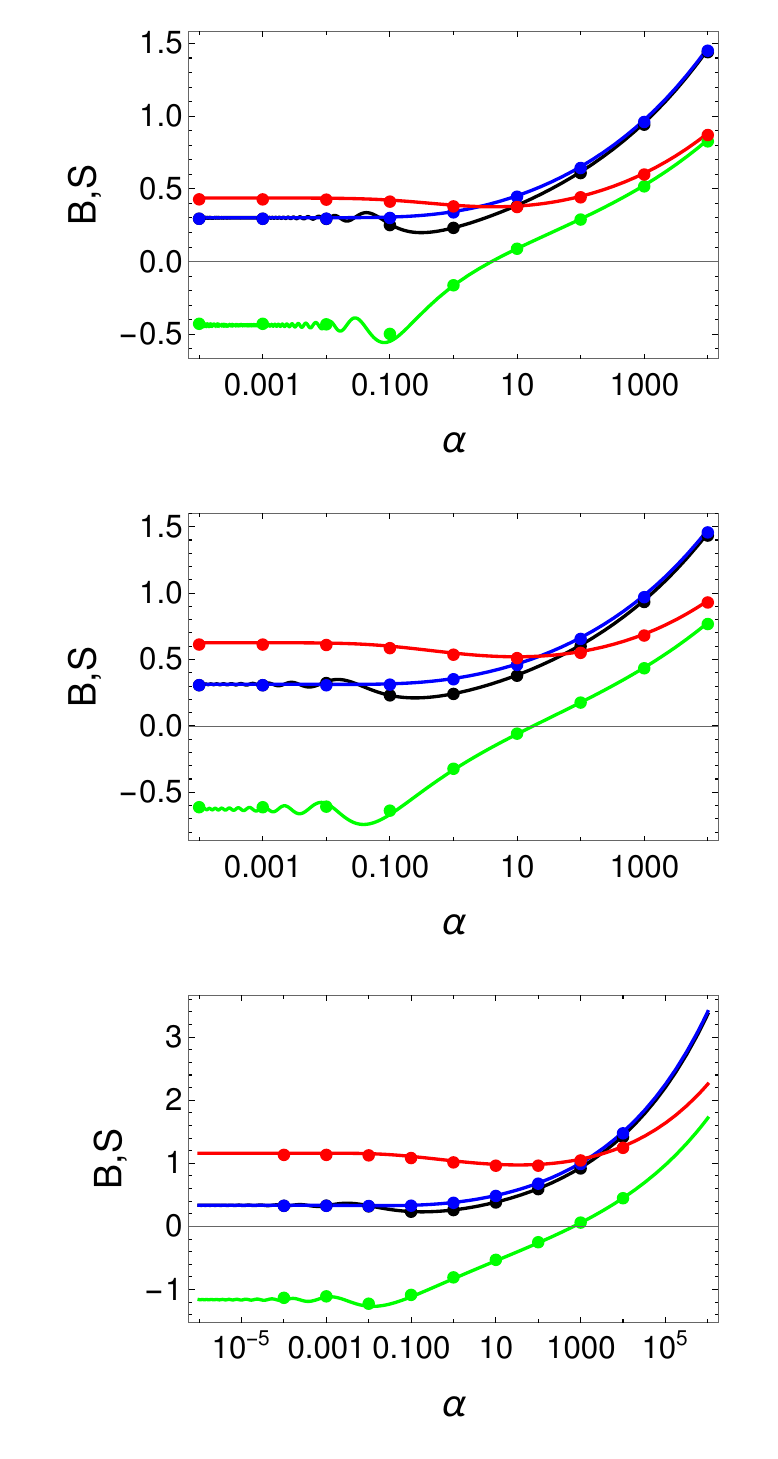}
    \caption{Broadening integrals $B(\alpha)$, $B(-\alpha)$ and shift integrals $S(\alpha)$, $- S(-\alpha)$ represented, respectively, by black, blue, green and red lines for MTVA-calculations and symbols for MBA-calculations as functions of the parameter $\alpha$ for Lennard-Jones interactions 12-6 (top panel), 12-5 (middle panel) and 12-4 (bottom panel).}
    \label{fig1}
\end{figure}

The results of numerical integrations for the trajectory integrals corresponding to Lennard-Jones interactions 12-6, 12-5 and 12-4 are plotted in Figure~\ref{fig1} for possible combinations of $\Delta C_{12}^{\prime}$ and $\Delta C_{m}^{\prime}$ signs. As can be seen from Fig.~\ref{fig1} for $m=6$, the trajectory-integrals dependence on $\alpha$ corresponds to the curves reported by Hindmarch et al. \cite{Hindmarsh1967}, and the two other panels ($m=5$ and 4) show similar behaviour but with different asymptotic values for $\alpha \rightarrow 0$ and smaller $\alpha$ values needed to visualize the convergence. (Remember that the definitions of $\alpha$ and the integration variable $x$ differ for different $m$.)

\subsubsection{Temperature dependence}

With the help of trajectory integrals, the linewidth $\gamma$ and shift $\delta$ (both in s$^{-1}$) corresponding to the Lennard-Jones 12-6 potential are given by Eqs. (\ref{eq6}) and (\ref{eq7}).
However, in practice it is more convenient to work with broadening and shifting coefficients per unit pressure (atm). As $N = {N_{A}P}/{(RT)}$, where $N_{A}$ is Avogadro number, $P$ is pressure and $R$ is the universal gas constant, we can get from Eqs.~(\ref{eq6}) and (\ref{eq7}) the ratios $\gamma\left( \overline{\text{v}} \right)/P$ and $\delta\left( \overline{\text{v}} \right)/P$. Moreover, the commonly used wavenumber units of cm$^{-1}$ are preferable instead of frequency units s$^{-1}$, so that additional division by the speed of light $c$ should be performed. Finally, the temperature dependence contained in $N$ and in $\overline{v} = (8kT/\pi\mu)^{1/2}$ (where $k$ is Boltzmann constant and $\mu$ is the reduced mass) can be written down explicitly. We get the broadening and shifting coefficients (in cm$^{-1}$atm$^{-1}$) as functions of temperature:
\begin{equation}
\label{eq14}
\widetilde{\gamma}(T) = 1.6444 \cdot 10^{14}\left| \Delta C_{6}^{\prime} \right|^{\frac{2}{5}}\mu^{- 0.3}T^{- 0.7}B(\alpha(T)) \ ,
\end{equation}
\begin{equation}
\label{eq15}
\widetilde{\delta}(T) = 8.222 \cdot 10^{13}\left| \Delta C_{6}^{\prime} \right|^{\frac{2}{5}}\mu^{- 0.3}T^{- 0.7}S(\alpha(T)) \ ,
\end{equation}
where $\Delta C_{6}^{\prime}$ are in rad~s$^{-1}$cm$^{6}$, $\mu$ in Dalton and $T$ in K.

\begin{figure}[ht]
    \centering
    \begin{subfigure}[t]{0.48\textwidth}
        \centering
        \includegraphics[width=\linewidth]{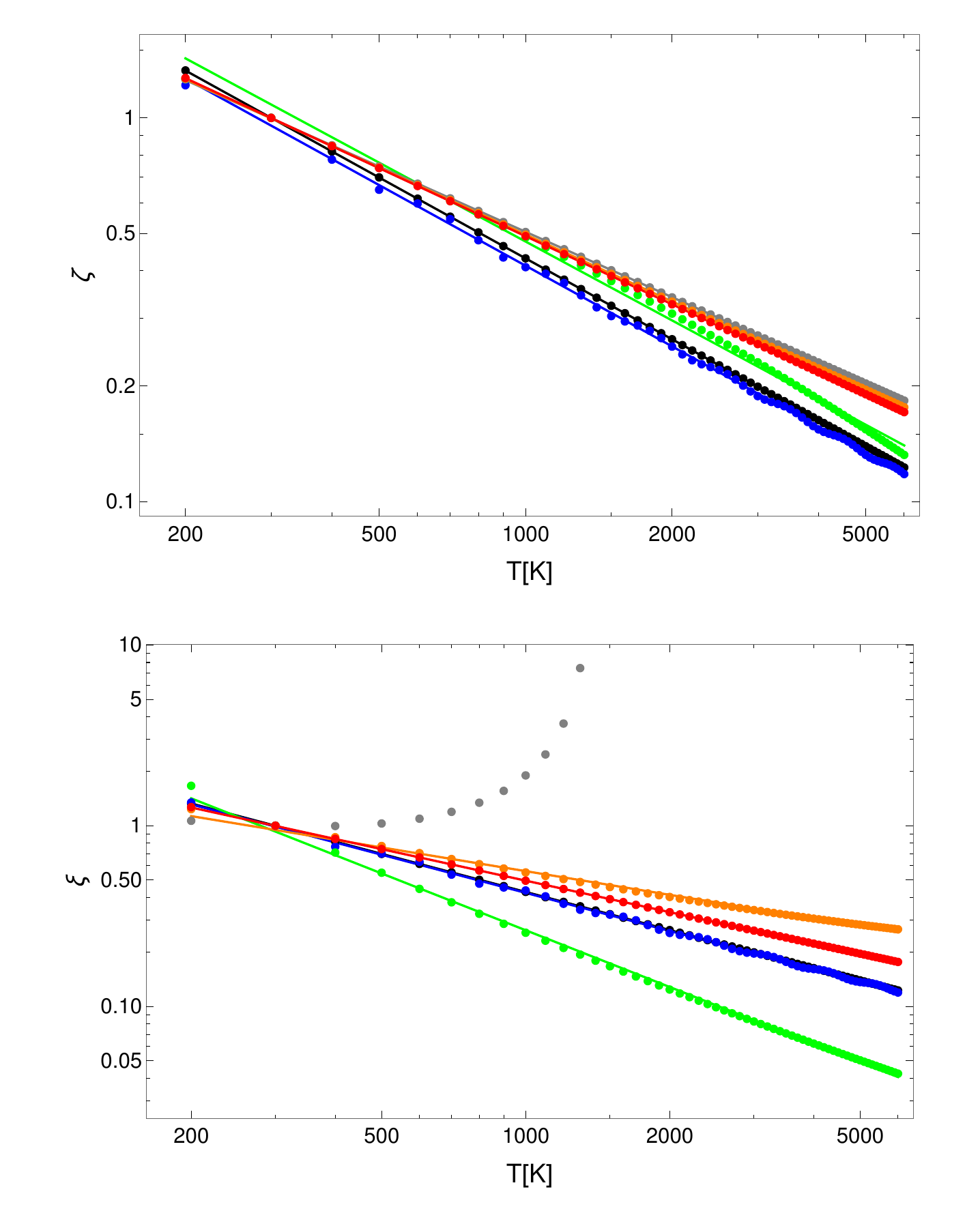}
        \label{fig2left}
    \end{subfigure}
    \hfill
    \begin{subfigure}[t]{0.48\textwidth}
        \centering
        \includegraphics[width=\linewidth]{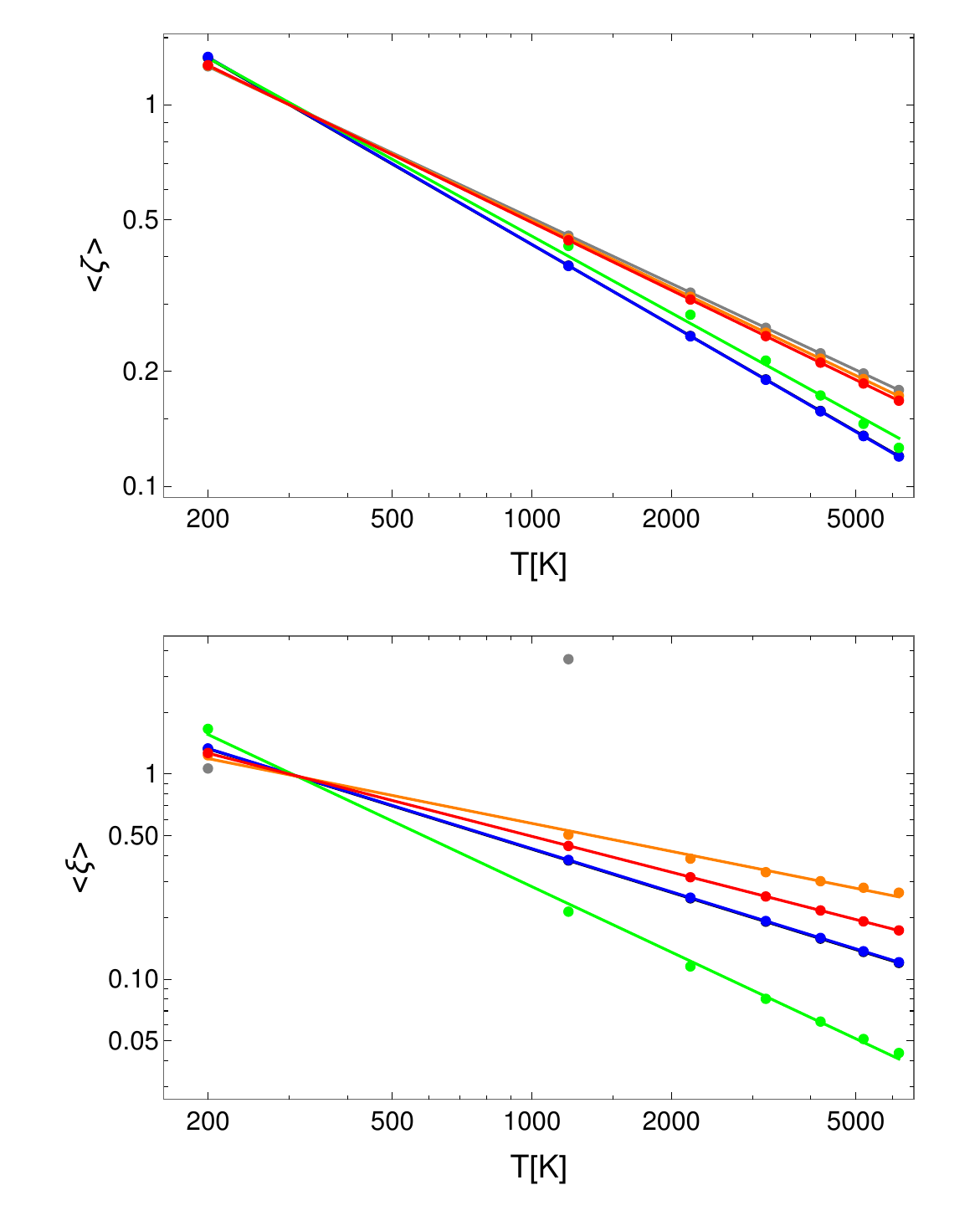}
        \label{fig2right}
    \end{subfigure}
    \caption{Reduced broadening (upper panels) and shifting (lower panels) coefficients as functions of temperature obtained with the MTVA (left panels) and MBA (right panels) for various $\alpha$-values (symbols) and their fits to one-power laws (solid lines). Plots corresponding to $\alpha~=~10^{-4}$, $10^{-2}$, $10^{0}$, $10^{1}$, $10^{2}$ and $10^{4}$ are depicted, respectively, in black, blue, green, gray, orange and red.}
    \label{fig2}
\end{figure}

Introducing a reference temperature $T_{\rm ref}$ and the associated $\widetilde{\gamma}\left( T_{\rm ref} \right)$, $\widetilde{\delta}(T_{\rm ref})$ coefficients allows analysis of the temperature dependence for various sets of Lennard-Jones parameters $\Delta C_{12}$, $\Delta C_{6}$ (i.e. various $\alpha$ values):
\begin{multline}
\zeta = \frac{\widetilde{\gamma}(T)}{\widetilde{\gamma}\left( T_{\rm ref} \right)} = \left( \frac{T}{T_{\rm ref}} \right)^{- 0.7}\frac{B(\alpha(T))}{B(\alpha(T_{\rm ref}))} \\
= \left( \frac{T}{T_{\rm ref}} \right)^{- 0.7}\frac{B(\alpha(T))}{B({(T_{\rm ref}/T)}^{0.6}\alpha(T))} \ ,
\label{eq16}
\end{multline}
\begin{multline}
\xi = \frac{\widetilde{\delta}(T)}{\widetilde{\delta}\left( T_{\rm ref} \right)} = \left( \frac{T}{T_{\rm ref}} \right)^{- 0.7}\frac{S(\alpha(T))}{S(\alpha(T_{\rm ref}))} \\
= \left( \frac{T}{T_{\rm ref}} \right)^{- 0.7}\frac{S(\alpha(T))}{S({(T_{\rm ref}/T)}^{0.6}\alpha(T))} \ ,
\label{eq17}
\end{multline}
where the definition of $\alpha$ leading to $$\alpha(T) = \alpha(T_{\rm ref})\left( \frac{T}{T_{\rm ref}} \right)^{\frac{3}{5}}$$ has been used. Note that our definitions of $\zeta$ and $\xi$ differ from those adopted by Cybulski et al. \cite{Cybulski2013}, who got dimensionless broadening and shifting coefficients by dividing $\widetilde{\gamma}$ and $\widetilde{\delta}$ by the factors standing before the products of $T^{-0.7}$ and the trajectory integrals. Small $\alpha$-values (dominant $C_{6}$ attraction) lead to unit ratios of trajectory integrals, so that the power laws
\begin{equation}
\label{eq18}
\widetilde{\gamma}(T) = \widetilde{\gamma}\left( T_{\rm ref} \right)\left( \frac{T}{T_{\rm ref}} \right)^{n}  \ ,
\end{equation}
\begin{equation}
\label{eq19}
\widetilde{\delta}(T) = \widetilde{\delta}\left( T_{\rm ref} \right)\left( \frac{T}{T_{\rm ref}} \right)^{q}
\end{equation}
with $n = q = -0.7$ are obtained in this limit case. The validity of this simple power law with respect to the sign of the line shift is discussed in detail below. To investigate the applicability of these simple power laws for increasing $\alpha$, we plotted in Figure~\ref{fig2} (left-hand panels) examples of the $T$-dependence of the reduced broadening and shifting coefficients (with $T_{\rm ref}~=~300~$K) for $\alpha~=~10^{-4}$, $10^{-2}$, $10^{0}$, $10^{1}$, $10^{2}$, $10^{4}$ over a wide temperature range between 200 and 6000~K and performed their linear fits (indicated by solid lines). For the $\alpha$-values listed, the powers $n~=~-0.70033(6)$, $-0.698(2)$, $-0.683(6)$, $-0.5658(2)$, $-0.5794(2)$, $-0.58941(2)$ were deduced for line broadening, confirming thus the $-0.7$ value for small $\alpha$ and the high-limit value $-0.59$ characteristic of the pure $r^{-12}$ interaction. For line shifts, the points corresponding to $\alpha~=~10$ do not follow a linear dependence in log-log coordinates, so that it was impossible to determine the corresponding power. For the remaining $\alpha$-values the powers were found to be $q~=~-0.7002(1)$, $-0.701(3)$, $-1.044(6)$, $-0.436(4)$, $-0.5791(2)$, again in agreement with the limit values $-0.7$ and $-0.59$ for $\alpha \rightarrow 0$ and $\alpha \rightarrow \infty$, respectively.

\begin{figure}[!ht]
    \centering
    \includegraphics[width=0.7\linewidth]{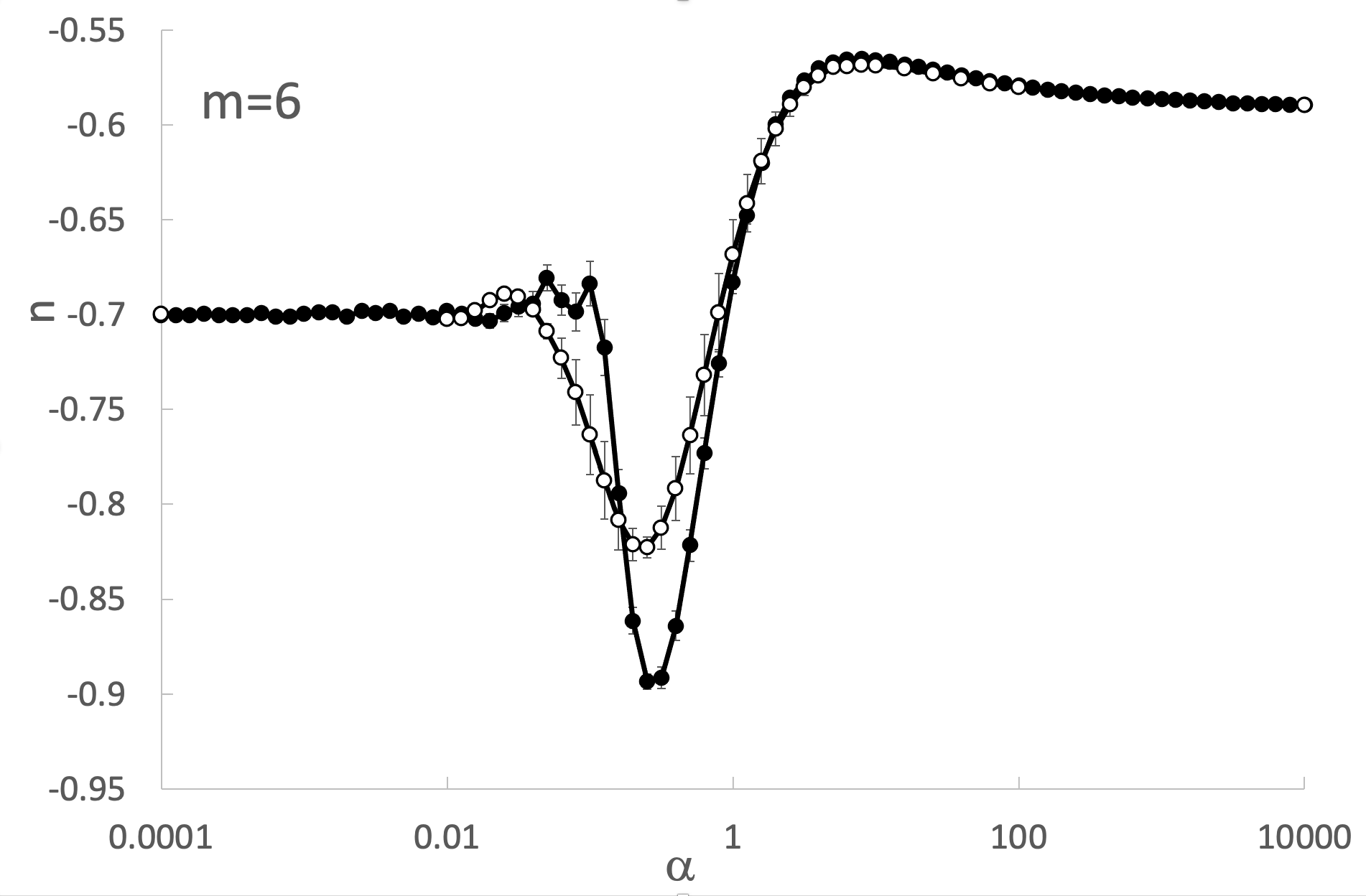}
    \includegraphics[width=\linewidth]{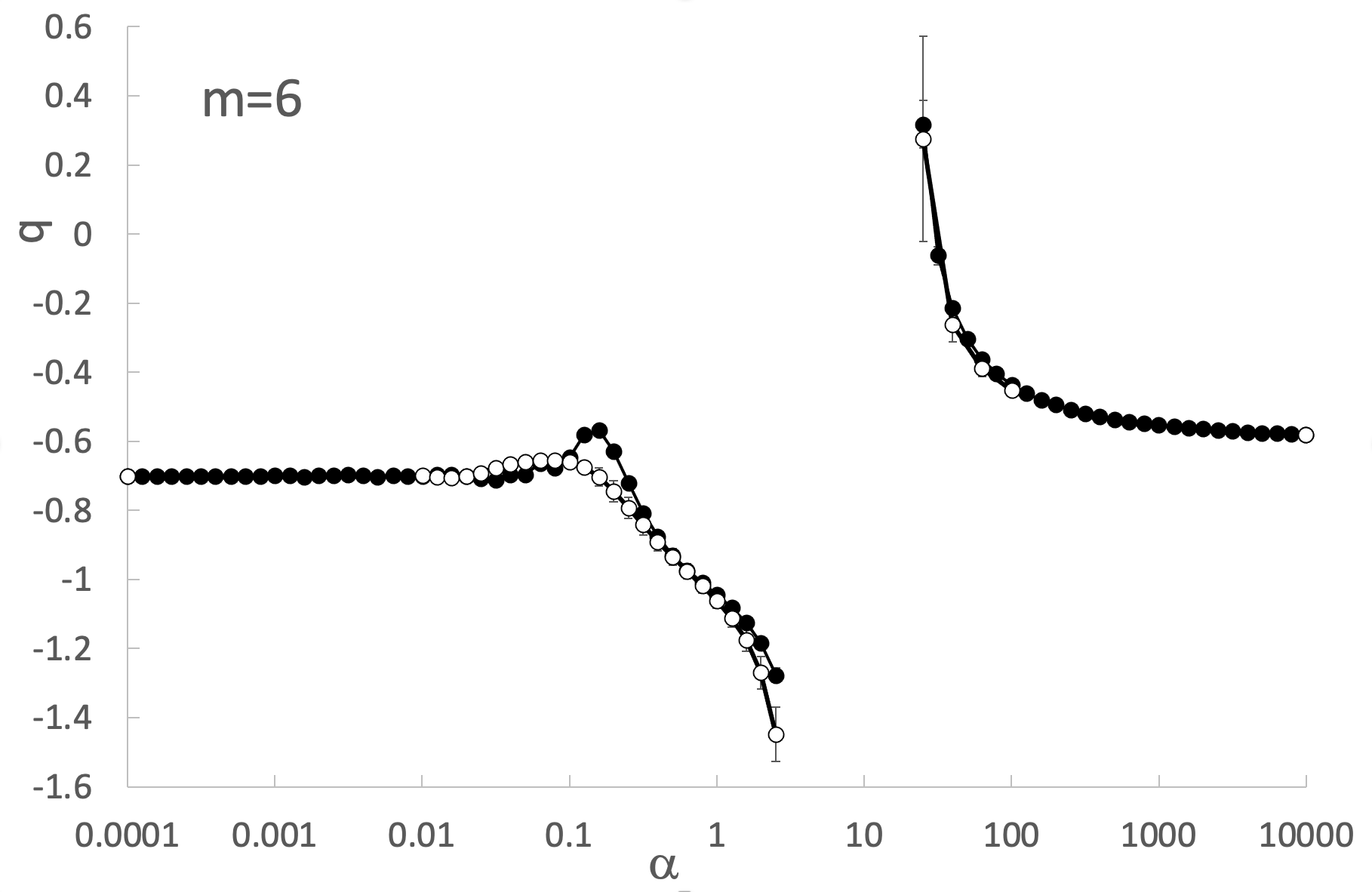}
    \caption{Powers extracted from linear fits in log-log coordinates as functions of $\alpha$ for line-broadening (upper panel) and shift (lower panel) coefficients.}
    \label{fig3}
\end{figure}

The validity of the power law is addressed in more detail in Figure~\ref{fig3}, where the powers extracted from linear fits in log-log coordinates are plotted versus $\alpha$. Both for linewidth and shift powers, the $\alpha$-dependence is like those reported by Cybulski et al. \cite{Cybulski2013}. The left- and right-hand sides of the lower-panel plot refer to negative- (red-) and positive- (blue-) shift domains, where the interactions are dominated, respectively, by attractive and repulsive forces. For intermolecular interactions corresponding to $\alpha$ about $10$ the shift-sign change occurs (the trajectory integral $S(\alpha)$ becomes equal to zero at $\alpha^{*} = 3.9792$), but the exact temperature for which it happens depends on the  molecular system considered through $\alpha$'s dependence on $\Delta C_{12}^{\prime}$ and $\Delta C_{6}^{\prime}$. In this $\alpha$ region, where a strong competition between attraction and repulsion takes place, as already pointed out by Cybulski et al. \cite{Cybulski2013}, the simple power law does not apply to the temperature dependence of line shifts (see also the case $\alpha = 10$ on the lower panel of Figure~\ref{fig2}).

\subsection{Maxwell-Boltzmann average over velocities}
Following  work by Mizushima \cite{Mizushima1951}, the velocity-averaged expressions for the linewidth and shift (in frequency units) are given by
\begin{equation}
\label{eq20}
\left\langle \gamma \right\rangle = (2\pi)^{- 1}\int_{}^{}{F(s)\left\lbrack 1 - \cos{\eta(s)} \right\rbrack ds} \ ,
\end{equation}
\begin{equation}
\label{eq21}
\left\langle \delta \right\rangle = (2\pi)^{- 1}\int_{}^{}{F(s)\sin{\eta(s)} ds} \ ,
\end{equation}
where the collision parameter $s = \left\{ b,v \right\}$ and
$$
F(s) ds = 8\pi^{1/2}\left( \frac{\mu}{2kT} \right)^{3/2}b db v^{3}e^{-\frac{\mu v^{2}}{2kT}}dv N.
$$
Like the MTVA calculations considered in the previous subsection, the integration on $b$ leads to trajectory integrals but now their parameter $\alpha$ is velocity-dependent and Maxwell-Boltzmann averaging should be performed. The natural choice for the dimensionless velocity v is such that $v = \text{v}v_{0}$ with the most probable velocity $v_{0} = \left( \frac{2kT}{\mu} \right)^{1/2}$. In terms of v, the new expression for $\alpha$ reads
\begin{equation}
\label{eq22}
\alpha_{MB} = \frac{7}{2^{7/5}3^{1/5}\pi^{6/5}}\left( \frac{2kT}{\mu} \right)^{3/5}\frac{\Delta C_{12}^{\prime}}{\left| \Delta C_{6}^{\prime} \right|^{11/5}}\text{v}^{6/5}
\end{equation}
and depends explicitly on the temperature. Eqs.~(\ref{eq20}) and (\ref{eq21}) become
\begin{multline}
\label{eq23}
\left\langle \gamma(T) \right\rangle = \frac{2^{\frac{9}{5}}3^{\frac{2}{5}}}{\pi^{\frac{1}{10}}}\left( \frac{2kT}{\mu} \right)^{\frac{3}{10}}\left| \Delta C_{6}^{\prime} \right|^{\frac{2}{5}}\int_{0}^{\infty}{\text{v}^{\frac{13}{5}}e^{- \text{v}^{\text{2}}}B(\alpha_{MB}(\text{v},T))d\text{v}}N \\
\equiv \frac{2^{\frac{9}{5}}3^{\frac{2}{5}}}{\pi^{\frac{1}{10}}}\left( \frac{2kT}{\mu} \right)^{\frac{3}{10}}\left| \Delta C_{6}^{\prime} \right|^{\frac{2}{5}}\left\langle B(T) \right\rangle N \ ,
\end{multline}
\begin{multline}
\label{eq24}
\left\langle \delta(T) \right\rangle = \frac{2^{\frac{4}{5}}3^{\frac{2}{5}}}{\pi^{\frac{1}{10}}}\left( \frac{2kT}{\mu} \right)^{\frac{3}{10}}\left| \Delta C_{6}^{\prime} \right|^{\frac{2}{5}}\int_{0}^{\infty}{\text{v}^{\frac{13}{5}}e^{- \text{v}^{\text{2}}}S(\alpha_{MB}(\text{v},T))d\text{v}}N \\
\equiv \frac{2^{\frac{4}{5}}3^{\frac{2}{5}}}{\pi^{\frac{1}{10}}}\left( \frac{2kT}{\mu} \right)^{\frac{3}{10}}\left| \Delta C_{6}^{\prime} \right|^{\frac{2}{5}}\left\langle S(T) \right\rangle N \ ,
\end{multline}
where the velocity-averaged trajectory integrals $\left\langle B(T) \right\rangle$ and $\left\langle S(T) \right\rangle$ keep parametric dependence on the interacting system via $\Delta C_{12}^{\prime}$, $\Delta C_{6}^{\prime}$ and $\mu$.

\subsubsection{Trajectory integrals}
To get an idea on the influence of velocity averaging on the trajectory integrals, we can rewrite $\left\langle \gamma(T) \right\rangle$, $\left\langle \delta(T) \right\rangle$ with the same factors $2\left( \frac{3\pi}{8} \right)^{2/5}\left| \Delta C_{6}^{\prime} \right|^{2/5}{\overline{v}}^{3/5}N$, $\left( \frac{3\pi}{8} \right)^{2/5}\left| \Delta C_{6}^{\prime} \right|^{2/5}{\overline{v}}^{3/5}N$ as in Eqs.~(\ref{eq6}) and (\ref{eq7}):
\begin{multline}
\label{eq25}
\left\langle \gamma(T) \right\rangle = 2\left( \frac{3\pi}{8} \right)^{2/5}\left| \Delta C_{6}^{\prime} \right|^{2/5}{\overline{v}}^{3/5}N \cdot 2^{\frac{7}{5}}\pi^{- \frac{1}{5}} \\
\times\int_{0}^{\infty}{\text{v}^{\frac{13}{5}}e^{- \text{v}^{\text{2}}}B(\alpha_{MB}(\text{v},T))d\text{v}} \ ,
\end{multline}
\begin{multline}
\label{eq26}
\left\langle \delta(T) \right\rangle = \left( \frac{3\pi}{8} \right)^{2/5}\left| \Delta C_{6}^{\prime} \right|^{2/5}{\overline{v}}^{3/5}N \cdot 2^{\frac{7}{5}}\pi^{- \frac{1}{5}} \\
\times \int_{0}^{\infty}{\text{v}^{\frac{13}{5}}e^{- \text{v}^{\text{2}}}S(\alpha_{MB}(\text{v},T))d\text{v}} \ .
\end{multline}
On the other hand, $\alpha_{MB}$ can be expressed in terms of $\alpha$:
\begin{multline}
\label{eq27}
\alpha_{MB} = \frac{7}{2^{7/5}3^{1/5}\pi^{6/5}}\left( \frac{2kT}{\mu} \right)^{3/5}\frac{\Delta C_{12}^{\prime}}{\left| \Delta C_{6}^{\prime} \right|^{11/5}}\text{v}^{6/5} \\
= \frac{7}{2^{7/5}3^{1/5}\pi^{6/5}}\left( \frac{\pi}{4} \right)^{3/5}{\overline{v}}^{6/5}\frac{\Delta C_{12}^{\prime}}{\left| \Delta C_{6}^{\prime} \right|^{11/5}}\text{v}^{6/5} = \left( \frac{\pi}{4} \right)^{3/5}\alpha\text{v}^{6/5} \ .
\end{multline}
So, the quantities $2^{\frac{7}{5}}\pi^{- \frac{1}{5}}\left\langle B(\alpha_{MB}(\alpha)) \right\rangle$ and $2^{\frac{7}{5}}\pi^{- \frac{1}{5}}\left\langle S\left( \alpha_{MB}(\alpha) \right) \right\rangle$ can be plotted as functions of $\alpha$ and compared to $B(\alpha)$ and $S(\alpha)$, respectively. Because of the double integration required and consequently increased CPU time, computations were performed for some fixed $\alpha$-values (see symbols in Fig.~\ref{fig1}); these results clearly show only negligible differences with respect to the MTVA-results. The most pronounced differences are seen for the line-shift integrals $S(\alpha)$ at $\alpha = 0.1$, where the oscillations observed for MTVA data are reduced. (This fact was also mentioned for broadening by Cybulski et al. \cite{Cybulski2013}).

\subsubsection{Temperature dependence}
The MB-averaged broadening and shifting coefficients are given by
\begin{equation}
\label{eq28}
\left\langle \widetilde{\gamma}(T) \right\rangle = 3.4516 \cdot 10^{14}\left| \Delta C_{6}^{\prime} \right|^{0.4}\mu^{- 0.3}T^{- 0.7}\left\langle B(T) \right\rangle \ ,
\end{equation}
\begin{equation}
\label{eq29}
\left\langle \widetilde{\delta}(T) \right\rangle = 1.7258 \cdot 10^{14}\left| \Delta C_{6}^{\prime} \right|^{\frac{2}{5}}\mu^{- 0.3}T^{- 0.7}\left\langle S(\text{T}) \right\rangle \ ,
\end{equation}
where, as previously (see Eqs.~(\ref{eq14}) and (\ref{eq15})), \(\Delta C_{6}^{\prime}\) are in rad~s$^{-1}$cm$^{6}$, $\mu$ in Dalton and $T$ in K. Introducing the MBA-equivalents of Eqs.~(\ref{eq16}) and (\ref{eq17})
\begin{multline}
\label{eq30}
\left\langle \zeta \right\rangle = \frac{\left\langle \widetilde{\gamma}(T) \right\rangle}{\left\langle \widetilde{\gamma}\left( T_{\rm ref} \right) \right\rangle} \\ = \left( \frac{T}{T_{\rm ref}} \right)^{- 0.7}\frac{\left\langle B(T) \right\rangle}{\left\langle B(T_{\rm ref}) \right\rangle} = \left( \frac{T}{T_{\rm ref}} \right)^{- 0.7}\frac{\left\langle B(\alpha_{MB}(T)) \right\rangle}{\left\langle B({(T_{\rm ref}/T)}^{0.6}\alpha_{MB}(T)) \right\rangle} \ ,
\end{multline}
\begin{multline}
\label{eq31}
\left\langle \xi \right\rangle = \frac{\left\langle \widetilde{\delta}(T) \right\rangle}{\left\langle \widetilde{\delta}\left( T_{\rm ref} \right) \right\rangle} = \left( \frac{T}{T_{\rm ref}} \right)^{- 0.7}\frac{\left\langle S(T) \right\rangle}{\left\langle S(T_{\rm ref}) \right\rangle} \\
= \left( \frac{T}{T_{\rm ref}} \right)^{- 0.7}\frac{\left\langle S(\alpha_{MB}(T)) \right\rangle}{\left\langle S({(T_{\rm ref}/T)}^{0.6}\alpha_{MB}(T)) \right\rangle} \ ,
\end{multline}
we can plot them against temperature for the same $\alpha$-values as in the MTVA case (right panels in Fig.~\ref{fig2}) and perform new fits in log-log coordinates to deduce the temperature-dependence exponents. As an additional integration on velocity is now needed, the CPU cost increases significantly, and only 7 temperature points are shown. Very similar values with respect to those of Sec.~3.1.2 are obtained for the temperature exponents of broadening and shifting: $n_{MB}~=~-0.7000040(4)$, $-0.7022(5)$, $-0.69(2)$, $-0.5687(4)$, $-0.5803(6)$, $-0.58954(8)$ and $q_{MB}~=~-0.69999(2)$, $-0.6987(1)$, $-1.06(2)$, $-0.45(1)$, $-0.5801(7)$. The number of points used for the fits was different in the MTVA and MBA cases, so only leading significant digits in the temperature exponents can reliably be compared. If we limit the comparison to two decimals, $n$ and $n_{MB}$ are identical for all $\alpha$ values considered except for $\alpha = 1$ (for which $n~=~-0.68$ and $n_{MB}~=~-0.66$) whereas $q$ and $q_{MB}$ differ for $\alpha = 1$ ($q~=~-1.04$, $q_{MB}~=~-1.06$) and for $\alpha = 10^{2}$ ($q~=~-0.44$, $q_{MB}~=~-0.45$).

The corresponding one-power laws
\begin{equation}
\label{eq32}
\left\langle \widetilde{\gamma}(T) \right\rangle = \left\langle \widetilde{\gamma}\left( T_{\rm ref} \right) \right\rangle\left( \frac{T}{T_{\rm ref}} \right)^{ n_{MB}} \ ,
\end{equation}
\begin{equation}
\label{eq33}
\left\langle \widetilde{\delta}(T) \right\rangle = \left\langle \widetilde{\delta}\left( T_{\rm ref} \right) \right\rangle\left( \frac{T}{T_{\rm ref}} \right)^{ q_{MB}}
\end{equation}
tested for a range of $\alpha$-values provide the temperature exponents represented by empty circles in Fig.~\ref{fig3}. In the case of line shifts (lower panel) the use of Maxwell-Boltzmann average does not significantly modify the results: just the amplitude of the local ``oscillation'' (local maximum) in the left-hand portion of the curve is slightly reduced. In contrast, for the linewidths (upper panel), the averaging influences in a non-negligible manner the temperature exponents in the region of $\alpha$ values around 0.1-0.5, so for molecular systems corresponding to such $\alpha$ values the MBA calculations are preferable.

\subsection{Phase-shift theory with straight-line trajectories: applications}

The general considerations presented above are supported below by calculations for some test systems. To choose them, we took into account the availability of experimental data up to high temperatures (see Table~\ref{Table1}) and representativity of various leading interactions. NO and OH with the permanent dipole moments differing by an order of magnitude (0.158~D \cite{Stogrun1966} and 1.668~D \cite{Meerts1973}, respectively) were chosen as active molecules whereas Ar and N$_2$ --- a rare-gas atom and a non-polar molecule with the quadrupole as its leading multipole --- were taken as perturbers. With these combinations we probe the behaviour of the active molecule with weak and strong dipoles and the role of dispersion/induction and electrostatic interactions.

\begin{figure}[!ht]
    \centering
    \includegraphics[width=0.7\linewidth]{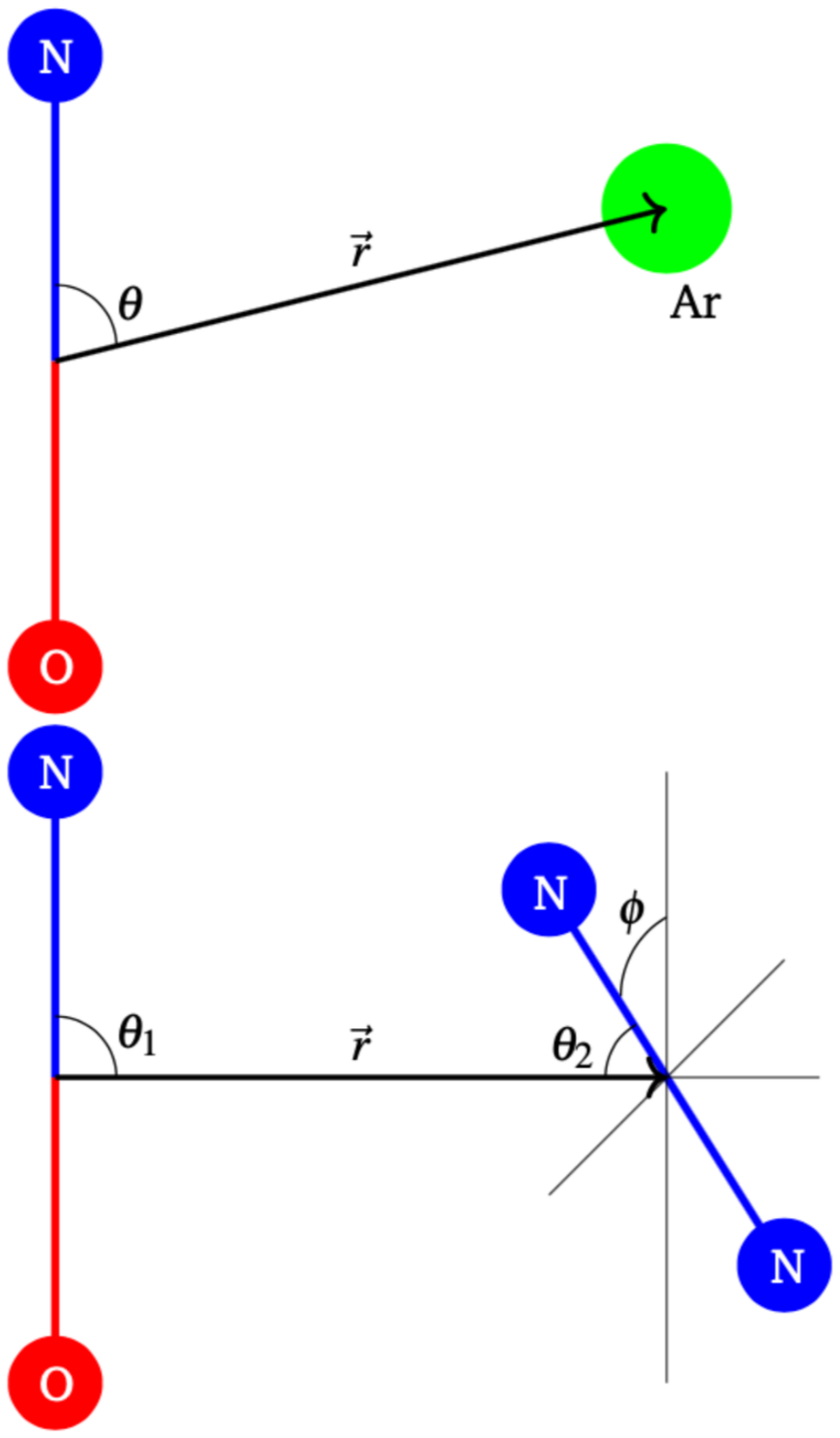}
    \caption{Collision geometry and Jacobi coordinates for NO perturbed by Ar and N$_2$.}
    \label{fig3supp}
\end{figure}

Inter-molecule potential energy surfaces (PES) of the complexes NO-Ar, NO-N$_2$, OH-Ar and OH-N$_2$ were computed \textit{ab initio} using the MOLPRO quantum chemistry package \cite{MOLPRO2020}  at  the  coupled-cluster level of theory CCSD(T):  RCCSD(T)/aug-cc-pV(X+d)z (spin-restricted) or UCCSD(T)/aug-cc-pV(X+d)z (spin-unrestricted), where X=T,Q.\citep{Knowles-RCCSD-1993} Since CCSD(T) is a single reference theory, the T1 diagnostic\citep{89LeTaxx.method} for all ground and excited state calculations was checked against both the 0.044 and 0.02 criteria  (criteria $\mathcal{C}_1$ and $\mathcal{C}_2$, herein) suggested by Rienstra-Kiracofe et al. \cite{Rienstra-Kiracofe2000} and by Lee and Taylor \cite{89LeTaxx.method}, respectively,  where T1 values larger than this indicate the need for a multireference electron correlation procedure. The corresponding inter-atomic distances were fixed at their equilibrium values, while the distance and orientation between the radiator (NO or OH) and perturber were varied over sets of grid points in Jacobi coordinates: $\lbrace r_i, \theta_j \rbrace$ for collisions with Ar or $\lbrace r_i, \theta_{1j}, \theta_{2k}, \phi_{l}\rbrace$ for collisions with N$_2$ (see Fig.~\ref{fig3supp}).

For collisions with Ar, the isotropic parts of the PESs corresponding to the ground and excited electronic states were extracted with the use of expansions over series of $l-$th rank Legendre polynomials $P_{l}$:
\[
V\left( r,\theta \right) = \sum_{l}{V_{l}\left( r \right)P_{l}(\cos\left( \theta \right))} \ ,
\]
where $V_{l}\left( r \right)$ are the radial potential terms with $l=0$ giving the isotropic component.
For collisions with N$_2$, the approach of limiting geometries for homonuclear-heteronuclear diatoms (A$_2$-BC) was used \cite{17BaCrPa} and the isotropic term was
\[
V(r)=\frac{1}{18} \lbrace 4V_H(r)+V_{L_1}(r)+V_{L_2}(r)+
2\lbrack V_{T_1}(r)+V_{T_2}(r)+2(V_{T_3}(r)+V_{X}(r)) \rbrack \rbrace \ ,
\]
where the indices on the potentials in the rhs indicate the contributing geometries (see   \cite{17BaCrPa} for more details).
The isotropic parts were further fitted by 12-6 Lennard-Jones expressions (see Eq.~\ref{eq4}).

We note here that within the phase-shift theory the accuracy of the final computed broadening parameters is limited to not the quality of \textit{ab initio} PESs but rather the theory itself. Phase-shift theory is an approximate frame whereby semi-classical methods for modelling the collision processes are made, such as the notion of a classical trajectory and the lack of rotation-vibration effects accounted for. Furthermore, we fit the isotropic interaction potential derived from our \textit{ab initio} PESs to a simple Lennard-Jones form, so any inaccuracies introduced by choice of quantum chemistry theory will not effect the computed broadening parameters with any appreciable magnitude since errors introduced by the Lennard-Jones approximation will be greater. What phase-shift theory allows us to do is derive values for the dimensionless parameter $\alpha$ and in turn the line broadening parameters for a general system of active molecule involved in an electronic transition and perturber.

\subsubsection{NO-Ar}
The interaction PESs of the NO-Ar complex were computed for the ground X\ $^2 \Pi$ and first excited A\ $^2 \Sigma^+$ electronic states of NO approximating the radiator by a rigid rotor with equilibrium bond lengths $r_{\rm e} = 1.15077$~\AA\ and $r_{\rm e} = 1.06434$~\AA, respectively.
Due to the computational costs of the RCCSD(T)/aug-cc-pV(X+d)z level of theory  (X=T,Q), a grid of 62 intermolecular distances NO--Ar corresponding to the range 2--15~\AA\ and only 5 angles between the inter-molecular distance vector and the NO molecular axis equal to 30, 60, 90, 120 and 150 degrees were considered, providing 310 points in total.

The T1 diagnostic never exceeded values of 0.017 and 0.032 for the ground- and excited-state calculations, respectively, both less than the $\mathcal{C}_1$ and $\mathcal{C}_2$ criteria. This confirms that our computed wavefunctions are described well by a single reference determinant and that there is minimal multireference character.



When introducing the Ar atom from non-linear geometries the X\ $^{2}\Pi$ ground state of NO is lowered in symmetry, creating a symmetric $A'$ and antisymmetric $A''$ states of the NO-Ar complex within the C$_{s}$ point group. This lowering in symmetry means that NO(A\ $^{2}\Sigma^{+}$)--Ar is no longer the lowest $A'$ symmetry state. An advantage of this system is that its weak interaction does not break the orthogonality of NO's molecular orbitals, meaning that one can converge a CCSD(T)/aug-cc-pV(X+d)z (X = T,Q) calculation to the excited NO(A)--Ar state through rotation of the electron orbitals centered on the active molecule. We rotated the outer electron orbital of NO obtained via an initial restricted Hartree-Fock (RHF) calculation which was then used in a second RHF calculation preceding the actual RCCSD(T) calculation. Testing was done using a CASSCF+icMRCI
(Complete Active Space Self-Consistent Field \cite{87Roos} + internally-contracted Multi-Reference Configuration Interaction\citep{11ShKnWe.methods})
approach and the aug-cc-pV(X+d)z (X=T,Q) basis sets, however we often obtained strange results where multiple discontinuities are seen over the potential minimum, especially within the excited-state PES. The RCCSD(T) calculations, however, produced more stable results with a lower dissociation energy, in accordance with the values predicted  by Holmes-Ross and Lawrance \cite{11HoLawxx}, Tsuji et al. \cite{94TsShiOb}, Alexander\cite{99Millard} and Sumiyoshi and Endo \cite{07SuEnxx}.

For the ground state's   isotropic part of 12-6 Lennard-Jones, an overall good fit was obtained, with a correct potential depth (maximal difference ``calculated'' $-$ ``fitted'' of 8~cm$^{-1}$) and a very reliable representation of the repulsive wall up to 3500~cm$^{-1}$ (5000~K). For the excited state, the shape of the computed potential differed strongly from a 12--6 Lennard-Jones one. Having in mind high-temperature applications, we performed a first fit (denoted FIT I below) for intermolecular distances corresponding to the repulsive-wall region; this choice allowed a good representation for energies $>$2000~cm$^{-1}$ and a nearly correct representation for energies between 1000 and 2000~cm$^{-1}$, but much worse results elsewhere. A second fit (FIT II) was performed on the attraction region, with a reliable potential representation beyond 5.4~\AA.

The results of the NO-Ar PESs fits using 12-6 Lennard-Jones expressions are collected in Table~\ref{Table2} together with broadening and shifting coefficients values calculated at given temperatures. Note that for the NO-Ar case the shift-sign-change condition $\alpha^*=3.9792$ is never realized with FIT I Lennard-Jones parameters (since $\alpha$ is always negative for our negative $\Delta C_{12}^{\prime}$-value); for FIT II the corresponding critical temperature $T^*$ (about $10^8$~K) is unreachable. First, it can be seen that the intermolecular distance region chosen for fits (repulsive wall for FIT I and ``pure-attraction'' zone for FIT II) influences strongly the resulting linewidth and shift. Much better agreement with measurements is obtained for the Lennard-Jones parameters of FIT I (underestimation by about 20\% or even less for $\widetilde{\delta}$ at 295~K), conversely FIT II leads to overestimates by nearly 100\%. A correct representation of the repulsive wall in the Lennard-Jones potential model therefore appears to be crucial for getting reliable line-shape parameters. This is easily understandable from the viewpoint of linewidths which are mainly due to short-range collisions but is rather surprising for line shifts which are produced for the most part by distant collisions.

\begin{table*}[!ht]
\scriptsize
    \centering
    \caption{NO-Ar Lennard-Jones interaction potential parameters for the initial ($i$) and final ($f$) electronic states, corresponding differences $\Delta C_{12}=C_{12f}-C_{12i}$, $\Delta C_6=C_{6f}-C_{6i}$ and other characteristics involved in line broadening and shifting calculations. Pairs of calculated line-shape coefficients (rounded to 3 decimals, as typical measurements) correspond to MTVA and MBA approaches. Measured coefficients are given as intervals when multiple data are available in the literature; values marked by an asterisk are calculated by the power law with reference-temperature values and temperatures exponents determined experimentally \cite{Chang1992}}
    \label{Table2}
    \begin{tabular}{lrrrr}
            \hline
            &NO-Ar (Fit I)&  &  NO-Ar (Fit II)&  \\
            \hline
            C$_{12i}$ (\text{cm}$^{-1}$ \text{\AA}$^{12}$)& 2 750 936 152&   &2 750 936 152& \\
            C$_{12f}$ (\text{cm}$^{-1}$ \text{\AA}$^{12}$) &384 146 855   &   &170 568 593 745& \\
            C$_{6i}$ (\text{cm}$^{-1}$  \text{\AA}$^{6}$) &1 052 272     &   &1 052 272& \\
            C$_{6f}$ (\text{cm}$^{-1}$  \text{\AA}$^{6}$) &3 920         &   &12 238 050& \\
            $\Delta C_{12}$ (10$^{9}$ \text{cm}$^{-1}$ \text{\AA}$^{12}$)&-2.36679& &167.818& \\
            $\Delta C_{6}$ (10$^{6}$ \text{cm}$^{-1}$  \text{\AA}$^{6}$)&-1.04835&&11.1858& \\
            $T$(K) &295&2800&295&2800 \\
           $\overline{v}$ (10$^{4}$ cm$^{-1}$) & 6.03702 & 18.599 & 6.03702 &18.599 \\
            $\alpha$ &-0.00465231 &-0.0179502&0.00180467&0.00696304\\
            $\widetilde{\gamma}$ (\text{cm}$^{-1}$ atm$^{-1}$)&0.206/0.202(calc)&0.043/0.042 (calc)&0.537/0.519 (calc)&0.111/0.107 (calc) \\
            &[0.25,0.27] (expt)&0.058 (expt*)&[0.25,0.27] (expt)&0.058 (expt*) \\
            $\widetilde{\delta}$ (\text{cm}$^{-1}$ atm$^{-1}$)&-0.149/-0.146 (calc)&-0.031/-0.030 (calc)&-0.383/-0.378  (calc)&-0.077/-0.078 (calc) \\
            & -0.16 (expt)&-0.043 (expt*)&-0.16 (expt)&-0.043 (expt*) \\
            \hline
        \end{tabular}
\end{table*}

\subsubsection{NO-N$_2$}

Similar to the NO-Ar case, the ground- and first-excited-state PESs of the NO-N$_2$ complex were calculated at the RCCSD(T)/cc-pV(Q+d)z level  of theory using MOLPRO \cite{MOLPRO2020}.
The T1 diagnostic never exceeded 0.022 and 0.017, respectively, both less than the 0.044 criterion \cite{Rienstra-Kiracofe2000} for all ground- and excited-state calculations, confirming minimal multireference character.

Seven leading configurations of NO--N$_2$ geometry were taken into account: two linear L, three perpendicular T, a parallel H, and the X configurations. As for NO-Ar, the near linear geometries of the excited state were the most difficult to converge and the  UCCSD(T)/aug-cc-pV(X+d)z (X = T, Q) level of theory was used.


The results of isotropic PESs fits to 12--6 Lennard-Jones expressions (a full-range fit for the ground state and a repulsive-wall fit for the excited state) are collected in Table~\ref{Table3} together with the corresponding broadening and shifting coefficients values calculated at different  temperatures. In comparison with the NO-Ar results, the calculated line-broadening coefficients are below measurements by approximately 30\% (instead of 20\%) and the underestimatation of the shifts at the high temperature of 2700~K rises to 44\%. Therefore, the case of perturbation by N$_2$ is less well described by the traditional phase-shift theory. We can evoke a strong sensibility of the calculated values to the Lennard-Jones-form fits, so that lowering by 10\% is not very meaningful. On the other hand, the repulsion is much stronger in the excited state for NO-N$_2$ than for NO-Ar (see $C_{12\ f}$ parameters in Tables~\ref{Table2} and \ref{Table3}) but the straight-line trajectories ignore this fact.

\begin{table}[!ht]
\footnotesize
    \centering
    \caption{Same as Table~\ref{Table2} but for NO-N$_2$}
    \label{Table3}
    \begin{tabular}{lrr}
            \hline
            C$_{12i}$ (\text{cm}$^{-1}$ \text{\AA}$^{12}$)& 2 153 076 015 &    \\
            C$_{12f}$ (\text{cm}$^{-1}$ \text{\AA}$^{12}$) &1 715 292 473 &   \\
            C$_{6i}$ (\text{cm}$^{-1}$  \text{\AA}$^{6}$)  &    827 259   &   \\
            C$_{6f}$ (\text{cm}$^{-1}$  \text{\AA}$^{6}$)  & 116 870      &   \\
            $\Delta C_{12}$ (10$^{9}$ \text{cm}$^{-1}$ \text{\AA}$^{12}$)& -0.43778 & \\
            $\Delta C_{6}$ (10$^{6}$ \text{cm}$^{-1}$  \text{\AA}$^{6}$)& 0.71039 & \\
            $T$(K) &295&2800\\
           $\overline{v}$ (10$^{4}$ cm$^{-1}$) & 6.56549 & 19.8627 \\
            $\alpha$ &-0.00224041 &-0.00845771 \\
            $\widetilde{\gamma}$ (\text{cm}$^{-1}$ atm$^{-1}$)&0.186/0.181(calc)&0.039/0.039 (calc)\\
            & [0.28,0.35] (expt)&0.056 (expt*)\\
            $\widetilde{\delta}$ (\text{cm}$^{-1}$ atm$^{-1}$)&-0.135/-0.132 (calc)&-0.029/-0.028 (calc) \\
            & [-0.17,-0.18] (expt)&-0.052 (expt*)\\
            \hline
        \end{tabular}
\end{table}

\subsubsection{OH-Ar}

For all geometries of the OH(X)--Ar complex, the   UCCSD(T)/aug-cc-pVQz  (ground electronic state) and   MRCI/aug-cc-pV5z (excited electronic state) levels of theory  were used. The T1 diagnostic (both for the ground and electronic states) on average gave 0.004 and only 9 points of which exceeded the 0.044 criterion \cite{Rienstra-Kiracofe2000} at a value of 0.075. For the first electronically excited state A$^2 \Sigma^+$, electronic orbitals were rotated as for NO-Ar. However, convergence to the correct energy was not reached, which was checked by studying the difference in ground and excited coupled cluster energies for the complex at a 15~\AA\ separation. A CASSCF+icMRCI approach was then used to compute the excited state PES using the larger aug-cc-pV5Z basis sets.
	
A grid of intermolecular distances corresponding to the range 2--15~\AA\ and the angles between the intermolecular distance vector and the OH molecular axis equal to 10--180 degrees in 10 degree steps provided 1275 points in total.
The ground and first excited electronic states of OH are the same as NO, when introducing the Ar atom from non-linear geometries, we see again the ground X\ $^2\Pi$ state of OH lift degeneracy and energetically lowered, creating a symmetric $A’$ and antisymmetric $A''$ state of the OH-Ar complex within the C$_s$ point group.


Among various Lennard-Jones parameter sets obtained for different fitting regions we selected the values corresponding to a reliable fit of the repulsive wall up to 2000~cm$^{-1}$ for the ground state and to the range 550--2800~cm$^{-1}$ for the excited state. These values are given in Table~\ref{Table4} together with the corresponding line-broadening and shift coefficients.

\begin{table}[!ht]
\footnotesize
    \centering
    \caption{Same as Tables~\ref{Table2}  but for OH-Ar}
    \label{Table4}
    \begin{tabular}{lrr}
            \hline
            C$_{12i}$ (\text{cm}$^{-1}$ \text{\AA}$^{12}$)& 124 179 416.6 &    \\
            C$_{12f}$ (\text{cm}$^{-1}$ \text{\AA}$^{12}$) &1 110 474 162.3 &   \\
            C$_{6i}$ (\text{cm}$^{-1}$  \text{\AA}$^{6}$)  &    95554.75719   &   \\
            C$_{6f}$ (\text{cm}$^{-1}$  \text{\AA}$^{6}$)  & 377.7450683 &   \\
            $\Delta C_{12}$ (10$^{7}$ \text{cm}$^{-1}$ \text{\AA}$^{12}$)& -1.37053 & \\
            $\Delta C_{6}$ (10$^{4}$ \text{cm}$^{-1}$  \text{\AA}$^{6}$)& -9.5177 & \\
            $T$(K) &295&2000\\
           $\overline{v}$ (10$^{4}$ cm$^{-1}$) & 7.23557 & 18.8398 \\
            $\alpha$ &-0.00656329 &-0.0206941 \\
            $\widetilde{\gamma}$ (\text{cm}$^{-1}$ atm$^{-1}$)&0.088/0.086(calc)&0.023/0.023 (calc)\\
            & - (expt)&0.018 (expt*)\\
            $\widetilde{\delta}$ (\text{cm}$^{-1}$ atm$^{-1}$)&-0.064/-0.062 (calc)&-0.017/-0.016 (calc) \\
            & - (expt)& - (expt*)\\
            \hline
        \end{tabular}
\end{table}

\subsubsection{OH-N$_2$}

Unfortunately, all out attempts to produce adequate PESs for the OH-N$_2$ complex were unsuccessful. Initially, the same methodology used for the NO-N$_2$ system was tried  for OH-N$_2$, but using the UCCSD(T) level of theory. However, this failed to converge for the system's energy even at a separation of 15~\AA. For the EoM (Equation-of-Motion) coupled-cluster method, the wavefunctions, describing their difference to the RHF wavefunctions computed prior, would often diverge to values $>$100 after only 2-3 iterations. Attempts were also made using a RMP2 (second-order restricted M{\o}ller-Plesset) perturbation theory, but correct convergence in the excited state energy was not possible. Tests were made using the aug-cc-pVXz (X=T,Q,5,6,T+d,Q+d,5+d,6+d) basis sets\citep{89Dunning.ai,93WoDuxx.ai} for both UCCSD(T) and  RMP2 calculations. Finally, icMRCI calculations using molecular orbitals obtained from state-averaged CASSCF calculations
were tried for the system, but extremely long convergence times were seen, which is not practical for the number of geometries we need to compute. We leave the OH-N$_2$ system for a later study, where we aim to use the CFOUR software \cite{CFOUR} to employ further coupled cluster methods on this system.

\subsection{Numerical potentials for straight-line trajectories}

\begin{figure}[!ht]
    \centering
    \includegraphics[width=0.7\linewidth]{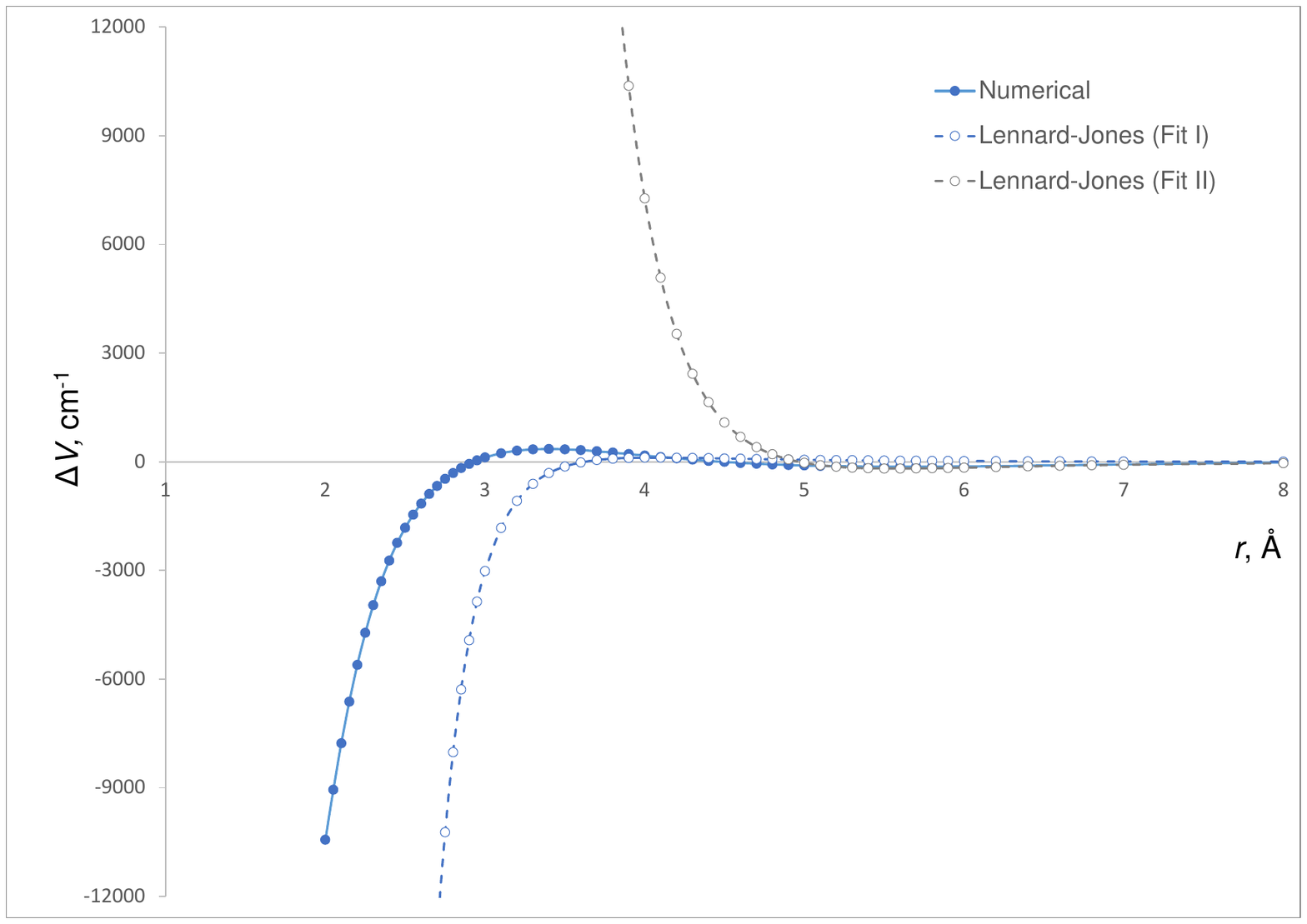}
    \caption{Differences of NO-Ar isotropic interactions in the final and initial states of the active molecule calculated from the numerical potential-energy surfaces (solid circles) and from Lennard-Jones fitted curves (Fit I and Fit II correspond to the excited-state dependency fitted on the repulsive-wall and attraction regions).}
    \label{fig4}
\end{figure}

\begin{figure}[!ht]
    \centering
    \includegraphics[width=\linewidth]{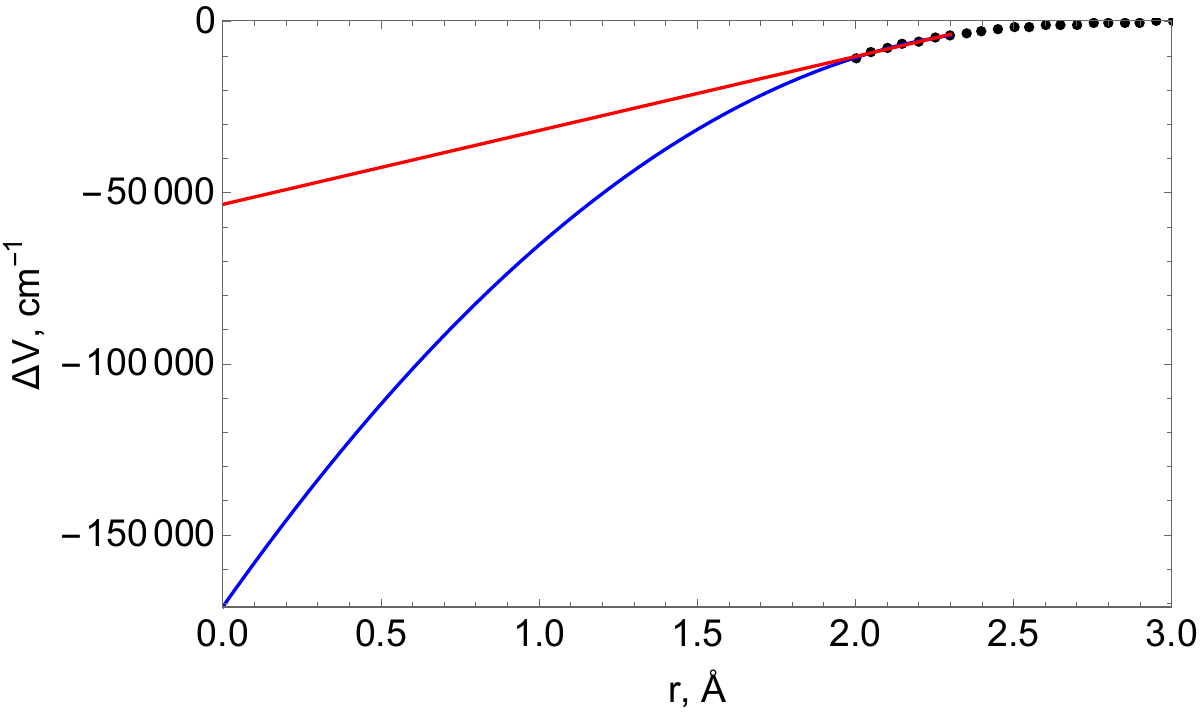}
    \includegraphics[width=\linewidth]{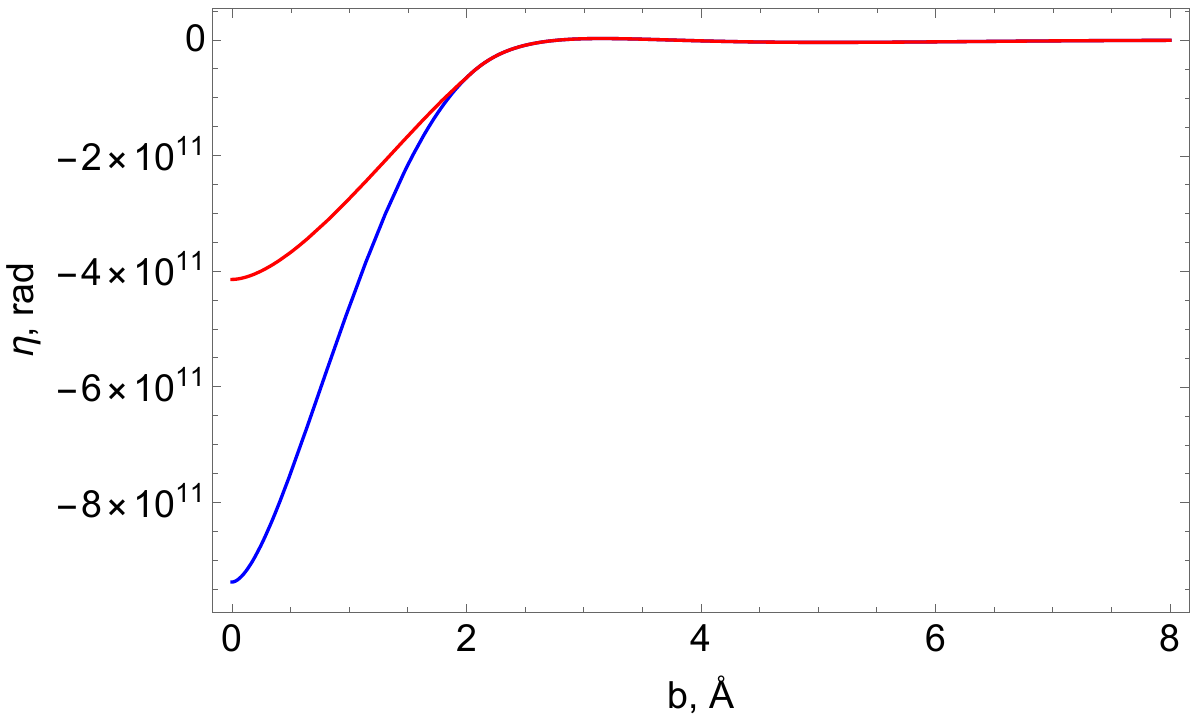}
    \caption{Extrapolations of $\Delta V$ to the region 0-2~\AA\ by polynomials of the first (red curve) and second (blue curve) order (upper panel) and corresponding dephasing functions (lower panel).}
    \label{fig5}
\end{figure}

The relatively straightforward formulae for linewidth/shift calculations (see Eqs.~(\ref{eq16}) and (\ref{eq17})) rely on a simplified representation of the isotropic intermolecular interaction by a 12-6 Lennard-Jones form, which enables analytic integration for the phase shift $\eta (b)$ and converged numerical integrations over the collision parameter(s) ($b$ in the MTVA and $b$ and $v$ in the MBA) for the linewidth $\gamma$ and shift $\delta$. However, as demonstrated, e.g., by our fits for the excited electronic state of NO interacting with Ar (Table~\ref{Table2}), the Lennard-Jones parameters are extremely sensitive to the choice of the intermolecular-distance interval and strongly influence the computed linewidth and shift. Therefore, in the frame of the MTVA adopted for simplicity, we attempted a numerical integration of the computed difference $\Delta V(r)=V_f (r)-V_i (r)$ of the isotropic interactions NO-Ar in the final and initial electronic states of the active molecule (shown in Figure~\ref{fig4} in comparison with Lennard-Jones-fit curves) to get the phase-shift dependence on $b$:
\begin{equation}
\label{eq35}
\eta(b)=\frac{2}{\overline{v}} \int_{0}^{\infty}\frac{\Delta V (r)}{\sqrt{1-(b/r)^2}}dr \ .
\end{equation}
Contrary to the inverse-power potential models where the unphysical zero value of the lower limit is accounted for by special $\Gamma$-functions \cite{Mizushima1951}, the use of a numerical $\Delta V$ (calculated in our NO-Ar case for the intermolecular distances from 2 to 15~\AA) gives rise to two problems: the need to extrapolate the calculated $\Delta V$ to the region 0--2~\AA\ and the divergence of $\Delta V$ at $r$ = 0.
 As extrapolations based on the full $r$-interval of computed $\Delta V$ gave unrealistic behaviour, extrapolation schemes built from regions near 2~\AA\ were tested. Two models — polynomial fits of first (PF1) and second (PF2) orders — issued from a ``nearly linear'' $\Delta V$-dependence region 2--2.3~\AA\ are considered below as examples. The corresponding extrapolated curves (Fig.~\ref{fig5}, upper panel) lead to different dephasing functions (Fig.~\ref{fig5}, lower panel) but result in practically identical line-broadening coefficients:
$\widetilde{\gamma}_{PF1}$ = $\widetilde{\gamma}_{PF2}~=~0.568$~cm$^{-1}$atm$^{-1}$ at 295~K and $\widetilde{\gamma}_{PF1}$=$\widetilde{\gamma}_{PF2}~=~0.184$~cm$^{-1}$atm$^{-1}$ at 2800~K. (Remember that in the Fourier-integral theory line broadening is determined by small $b$-values but cosine function oscillates strongly and cancels to a nearly zero net contributions from both $\eta_{PF1}$ and $\eta_{PF2}$.) The choice of extrapolation model appears therefore as not to be crucial for linewidths. However, the calculated values are overestimated with respect to measurements by a factor of two at both temperatures considered, and this discrepancy seems to indicate that the straight-line trajectory model is too rough for the true numerical potential. The line-shifting coefficients $\widetilde{\delta}_{PF1}=-0.0005$~cm$^{-1}$atm$^{-1}$ and
$\widetilde{\delta}_{PF2}=-0.0003$~cm$^{-1}$atm$^{-1}$ at 295~K as well as $\widetilde{\delta}_{PF1}=-0.00017$~cm$^{-1}$atm$^{-1}$ and
$\widetilde{\delta}_{PF2}=-0.00009$~cm$^{-1}$atm$^{-1}$ at 2800~K have correct negative signs but are by several orders of magnitude smaller than the measured values (see Table~\ref{Table1}). The shifts are mainly produced by distant collisions and their contributions, because of the sine function, can have opposite signs, leading in our case to nearly full compensation of positive and negative contributions. Various improvements of the trajectory model are examined below.

\section{Phase-shift theory with curved trajectories}
The model of straight-line trajectories used in the previous section assumes that the relative molecular motion is not at all influenced by molecular interactions (the isotropic potential is zero). This assumption is inconsistent not only with the key role of the isotropic potential in the broadening and shift formulae but also with the real motion of active and perturbing particles colliding with each other. Refinements of the model can be attempted considering that this is also the isotropic interaction potential which governs the relative molecular motion. For simplicity, MTVA frame ($v=\overline{v}$) will be assumed.

The use of the isotropic potential to govern the trajectory needs moreover to handle the fact that the interaction potential energies are different in the initial (usually, ground) and final (excited) electronic states of the radiator. While the energy of the absorbed photon excites the active molecule, the energy balance of the translational motion can be reasonably assumed to be not influenced by this change of the internal state. Indeed, this corresponds to the common classical-path approximation (decoupling of translational and internal degrees of freedom) used in IR/MW line-broadening theories. However, the relative motion is influenced since before ($t<0$) the (practically instantaneous) change of state (assumed to take place at $t=0$) the trajectory is driven by $V_i$ corresponding to the initial electronic state and after ($t>0$) this is done by $V_f$ related to the final state.

The phase shift given for rectilinear trajectories by Eq.~(\ref{eq4}) can now be re-written as ($p = 12$ or 6)
\begin{equation}
\label{eq36}
\eta_p=\Delta C_p^\prime\ \int_{-\infty}^{+\infty} r^{-p} (t) dt= \Delta C_p^\prime \lbrack \int_{-\infty}^0 r^{-p}_i (t) dt+\int_{0}^{+\infty} r^{-p}_f (t) dt \rbrack \ ,
\end{equation}
where the intermolecular distances $r_i$ and $r_f$ cover the respective half-trajectories. The time dependence of these distances can be further made explicit through a curved trajectory model. Two such models are considered below.

\subsection{Parabolic trajectories}
For parabolic trajectories the time dependence of the intermolecular distance is given by
\begin{equation}
\label{eq37}
\vec{r}(t)=\vec{r_c}+\vec{v_c}t+\frac{\vec{F_c}}{\mu} \frac{t^2}{2} \ .
\end{equation}
Here $\vec{r_c}$  denotes the distance of the closest approach for a given value of the impact parameter $b$, $\vec{v_c}$ stands for the relative velocity at this point and $\vec{F_c}$ represents the force derived from the isotropic potential:
\begin{equation}
\label{eq38}
\vec{F_c}=-\left(\frac{\partial V}{\partial r} \right)_{\vec{r}=\vec{r_c}} \frac{\vec{r_c}}{r_c} \ .
\end{equation}
The time dependence $r(t)$ of Eq.~(\ref{eq37}) can be re-written in a ``straight-line'' form
\begin{equation}
\label{eq39}
r(t)= \lbrack r_c^2+v_c^{\prime \ 2} t^2 \rbrack ^{1/2} \ ,
\end{equation}
where the apparent relative velocity $v_c^\prime$ is determined by ${v_c^\prime}^2=v_c^2+(\vec{F_c} \cdot \vec{r_c})/\mu$. This velocity can be related to the relative velocity before collision $v$ with the help of the angular momentum $v_cr_c=vb$ and energy-conservation $\mu v^2/2=\mu v_c^2/2+V(r_c)$ conditions as
\begin{equation}
\label{eq40}
v_c^\prime=v\left[1-V_i^\ast\left(r_c\right)-r_cV_i^{\ast\prime}\left(r_c\right)\ /2\right]^{1/2} \ ,
\end{equation}
where the dimensionless potential is given by $V^\ast\left(r\right)$ $\equiv2V(r)/(\mu v^2)$ and $V_i^{\ast\prime}\left(r_c\right)$ denotes the first derivative. Eq.~(\ref{eq40}) enables one to replace the integration on the impact parameter $b$ by integration on the more physical parameter $r_c$:
\begin{multline}
\label{eq41}
\int_{0}^{\infty}{\ldots b db\rightarrow \int_{r_{c\ min}}^{\infty}{\ldots r_c dr_c\left(\frac{v_c^\prime}{v}\right)^2}} \\=\int_{r_{c\ min}}^{\infty}{\ldots r_c dr_c\left[1-V_i^\ast\left(r_c\right)-r_c V_i^{\ast\prime}\left(r_c\right)\ /2\right]}
\end{multline}
starting at $r_{c\ min}$ corresponding to the head-on collision ($b=0$).

The energy conservation conditions ``before'' and ``after'' collision, respectively, $\mu v^2/2=\mu v_c^2/2+V_i(r_c)$ and $\mu{\widetilde{v}}^2/2=\mu v_c^2/2+V_f(r_c)$, where $\widetilde{v}$ is the relative velocity at $t\rightarrow\infty$, allow relating $\widetilde{v}$ to $v$:
\begin{equation}
\label{eq42}
\widetilde{v}=v\left[1-\frac{2\Delta V(r_c)}{\mu v^2}\right]^{1/2}	
\end{equation}
for each distance of the closest approach $r_c$ (each trajectory).

\subsubsection{Parabolic half-trajectories for Lennard-Jones isotropic potentials}
When the 12-6 Lennard-Jones form $V\left(r\right)$=$4\varepsilon\left[\left(\sigma/r\right)^{12}-\left(\sigma/r\right)^6\right]$ is assumed for mathematical convenience and in agreement with the leading term of the dispersion interactions, Eqs.~(\ref{eq38},\ref{eq40}-\ref{eq42}) become, respectively,
\begin{equation}
\label{eq43}
\vec{F_c}=\frac{24\varepsilon}{\sigma}\left[2\left(\frac{\sigma}{r}\right)^{13}-\left(\frac{\sigma}{r}\right)^7\right]\frac{\vec{r_c}}{r_c} \ ,
\end{equation}
\begin{equation}
\label{eq44}
v_c^\prime=v\left\{1+\frac{8\varepsilon}{\mu v^2}\left[5\left(\frac{\sigma}{r_c}\right)^{12}-2\left(\frac{\sigma}{r_c}\right)^6\right]\right\}^{1/2} \ , 					
\end{equation}
\begin{equation}
\label{eq45}
\int_{0}^{\infty}\ldots b d b\rightarrow\int_{r_{c\ min}}^{\infty} \ldots r_c dr_c\left\{1+\frac{8\varepsilon}{\mu v^2}\left[5\left(\frac{\sigma}{r_c}\right)^{12}-2\left(\frac{\sigma}{r_c}\right)^6\right]\right\}\ ,		
\end{equation}
\begin{equation}
\label{eq46}
\widetilde{v}=v\left\{1-\frac{8\varepsilon_i}{\mu v^2}\left[\left(\frac{\sigma_i}{r_c}\right)^{12}-\left(\frac{\sigma_i}{r_c}\right)^6\right]+\frac{8\varepsilon_f}{\mu v^2}\left[\left(\frac{\sigma_f}{r_c}\right)^{12}-\left(\frac{\sigma_f}{r_c}\right)^6\right]\right\}^{1/2} \ .
\end{equation}
Using Eqs.~(\ref{eq39}) and (\ref{eq44}), the first integral in Eq.~(\ref{eq36}) can be written as
\begin{multline}
\label{eq47}
\int_{-\infty}^{0}{r_i^{-p}(t)dt}=\frac{\sqrt\pi}{2}\frac{\Gamma\left(\frac{p-1}{2}\right)}{\Gamma\left(\frac{p}{2}\right)}r_c^{1-p}v^{-1} \\
\times \left\{1+\frac{8\varepsilon_i}{\mu v^2}\left[5\left(\frac{\sigma_i}{r_c}\right)^{12}-2\left(\frac{\sigma_i}{r_c}\right)^6\right]\right\}^{-1/2}
\end{multline}
and, with the additional use of Eq.~(\ref{eq46}), the second integral gives
\begin{multline}
\label{eq48}
\int_{0}^{\infty}{r_f^{-p}(t)dt}=\frac{\sqrt\pi}{2}\frac{\Gamma\left(\frac{p-1}{2}\right)}{\Gamma\left(\frac{p}{2}\right)}r_c^{1-p}v^{-1}
\left\{ 1-\frac{8\varepsilon_i}{\mu v^2}\left[\left(\frac{\sigma_i}{r_c}\right)^{12}-\left(\frac{\sigma_i}{r_c}\right)^6\right] \right. \\
\left. +\frac{8\varepsilon_f}{\mu v^2}\left[\left(\frac{\sigma_f}{r_c}\right)^{12}-\left(\frac{\sigma_f}{r_c}\right)^6\right]+\frac{8\varepsilon_f}{\mu v^2}\left[5\left(\frac{\sigma_f}{r_c}\right)^{12}-2\left(\frac{\sigma_f}{r_c}\right)^6\right] \right\} ^{-1/2} \ .
\end{multline}
Since $4\varepsilon\sigma^{12}=\hbar C_{12}^\prime$ and $4\varepsilon\sigma^6=\hbar C_6^\prime$, we get
\begin{multline}
\label{eq49}
\eta(r_c,v)=\left(\frac{63\pi\Delta C_{12}^\prime}{2\cdot 256 r_c^{11}}-\frac{3\pi\Delta C_6^\prime}{2\cdot 8r_c^5}\right)\frac{1}{v}
\left\{\left[1+\frac{2\hbar}{\mu v^2}\left(5\frac{C_{12i}^\prime}{r_c^{12}}-2\frac{C_{6i}^\prime}{r_c^6}\right)\right]^{-1/2} \right. \\
\left. +\left[1+\frac{2\hbar}{\mu v^2}\left(\frac{{\Delta C}_{12}^\prime}{r_c^{12}}-\frac{{\Delta C}_6^\prime}{r_c^6}\right)+\frac{2\hbar}{\mu v^2}\left(5\frac{C_{12f}^\prime}{r_c^{12}}-2\frac{C_{6f}^\prime}{r_c^6}\right)\right]^{-1/2}\right\} \ .
\end{multline}
Equations (\ref{eq1}) and (\ref{eq2}) for linewidth and shift are rewritten now in terms of $r_c$ ($v=\overline{v}$ in the MTVA frame):
\begin{equation}
\label{eq50}
\gamma=N\overline{v}\int_{r_{c\ min}}^{\infty}\left[1-\cos{\eta(r_c,\overline{v})}\right]\left[1+\frac{2\hbar}{\mu \overline{v}^2}\left(5\frac{C_{12i}^\prime}{r_c^{12}}-2\frac{C_{6i}^\prime}{r_c^6}\right)\right]r_c d r_c\ ,
\end{equation}
\begin{equation}
\label{eq51}
\delta=N\overline{v}\int_{r_{c\ min}}^{\infty}\sin{\eta(r_c,\overline{v})} \left[1+\frac{2\hbar}{\mu \overline{v}^2} \left(5\frac{C_{12i}^\prime}{r_c^{12}}-2\frac{C_{6i}^\prime}{r_c^6} \right) \right] r_c d r_c,
\end{equation}
with
\begin{multline}
\label{eq52}
r_{c\ min}=\sigma_i\left[\frac{2}{1+\sqrt{1+\mu \overline{v}^2/(2\varepsilon_i)}}\right]^{1/6} \\
=\left[\frac{C_{12i}^\prime}{C_{6i}^\prime}\frac{2}{1+\sqrt{1+2\mu v^2C_{12i}^\prime/(\hbar{(C_{6i}^\prime)}^2)}}\right]^{1/6} \ .
\end{multline}
Introducing further the dimensionless integration variable $y\equiv r_c/r_{c\ min}$ and the short-hand notations
\begin{multline}
\label{eq53}
\xi_1\equiv\frac{10\hbar C_{12i}^\prime}{\mu \overline{v}^2r_{c\ min}^{12}},\ \xi_2\equiv\frac{4\hbar C_{12f}^\prime}{\mu \overline{v}^2r_{c\ min}^{12}}, \ \xi_3\equiv\frac{63\pi{\Delta C}_{12}^\prime}{512 \overline{v}r_{c\ min}^{11}},\\
  \xi_4\equiv\frac{3\pi{\Delta C}_6^\prime}{16\overline{v}r_{c\ min}^5},\ \xi_5\equiv\frac{2\hbar({\Delta C}_{12}^\prime+5C_{12f}^\prime)}{\mu \overline{v}^2r_{c\ min}^{12}},\ \xi_6\equiv\frac{2\hbar({\Delta C}_6^\prime+2C_{6f}^\prime)}{\mu \overline{v}^2r_{c\ min}^6},	
\end{multline}
as well as switching to the broadening/shift coefficients, we arrive at (for $r_{c\ min}$ in \AA)
\begin{multline}
\label{eq54}
\widetilde{\gamma}=0.3561617 r_{c\ min}^2\mu^{-0.5}T^{-0.5} \\
\times \int_{1}^{\infty} \left[1-\cos \mathcal{A}(\xi_1,\xi_2,\xi_3,\xi_4,\xi_5,\xi_6,y) \right]\mathcal{B}(\xi_1,\xi_2,y)dy \ ,
\end{multline}
\begin{multline}
\label{eq55}
\widetilde{\delta}=0.3561617 r_{c\ min}^2\mu^{-0.5}T^{-0.5} \\
\times \int_{1}^{\infty} \sin \mathcal{A}(\xi_1,\xi_2,\xi_3,\xi_4,\xi_5,\xi_6,y) \mathcal{B}(\xi_1,\xi_2,y)dy
\end{multline}
with
\begin{multline}
\label{eq56}
\mathcal{A}\left(\xi_1,\xi_2,\xi_3,\xi_4,\xi_5,\xi_6,y\right)=\left[\xi_3y^{-11}-\xi_4y^{-5}\right]  \\ \times \left[\left(1+\xi_1y^{-12}-\xi_2y^{-6}\right)^{-0.5}  +\left(1+\xi_5y^{-12}-\xi_6y^{-6}\right)^{-0.5}\right], \\
\mathcal{B}\left(\xi_1,\xi_2,y\right)=y+\xi_1y^{-11}-\xi_2y^{-5}.
\end{multline}

\subsubsection{Parabolic half-trajectories for numerical potential difference}
Keeping in mind that the use of Lennard-Jones representations for the isotropic potentials in the initial and final states is approximate and extremely sensitive to the fit region model, we can also consider, as  done in Sec. 2.4, a numerically calculated $\Delta V$. Note that our numerical $\Delta V$ values are expressed in cm$^{-1}$ and should be multiplied by $2\pi c$ to be used in Eq.~(\ref{eq57}). The equivalent of Eq.~(\ref{eq36}) then reads
\begin{equation}
\label{eq57}
\eta=\int_{-\infty}^{0}\Delta V(r_i(t))dt+
\int_{0}^{+\infty}\Delta V(r_f(t))dt
\end{equation}
but the intermolecular distance and time are not more related by an analytical formula. Therefore, a numerical integration over $t$ should be performed by calculating sets of $r(t)$ and then sets of $\Delta V(r\left(t\right))$, for each half-trajectory (a grid of $r_c$ starting from $r_c\ min$ given by Eq.~(\ref{eq52})). For the half-trajectory driven by $V_i$  $r_i(t)$ is calculated via Eqs.~(\ref{eq39}) and (\ref{eq44}):
\begin{equation}
\label{eq58}
    r_i(t)=\left(r_c^2+\overline{v}^2\left\{1+\frac{8\varepsilon_i}{\mu \overline{v}^2}\left[5\left(\frac{\sigma_i}{r_c}\right)^{12}-2\left(\frac{\sigma_i}{r_c}\right)^6\right]\right\}t^2\right)^{1/2}
\end{equation}
with the term $\frac{8\varepsilon}{\mu \overline{v}^2}\left[\ldots\right]$ which can be further detailed as $1.13004 T^{-1}\left[5C_{12i}r_c^{-12}-2C_{6i}r_c^{-6}\right]$, where $T$ is in Kelvin, $C_{12i}$ is in cm$^{-1}$\AA$^{12}$ and $C_{6i}$ is in cm$^{-1}$\AA$^{6}$. For the second half-trajectory the role of $\overline{v}$ is played by $\widetilde{v}$ (see Eq.~(\ref{eq46})) which can also be rewritten as
\begin{equation}
\label{eq59}
    \widetilde{v}=\overline{v} \left[1+\ 1.13004 T^{-1}\left(\Delta C_{12}r_c^{-12}-\Delta C_6r_c^{-6}\right)\right]^{1/2} \ ,	
\end{equation}
and the intermolecular distance is given by
\begin{equation}
\label{eq60}
r_f\left(t\right)=\left\{ r_c^2+\widetilde{v}^2 \left[ 1+\frac{1.13004 T^{-1}(5C_{12i}r_c^{-12}-2C_{6i}r_c^{-6})}{1+1.13004 T^{-1}\left(\Delta C_{12}r_c^{-12}-\Delta C_6r_c^{-6}\right)}\right]t^2\right\}^{1/2}. 	
\end{equation}
The resulting broadening and shift coefficients are determined by
\begin{multline}
\label{eq61}
    \widetilde{\gamma}=0.356162\mu^{-0.5}T^{-0.5}\int_{r_{c\ min}}^{\infty}\left[1-\cos{\eta(r_c,\overline{v})}\right] \\
    \times \left[1+1.13004 T^{-1}\left(5C_{12\ i}r_c^{-12}-2C_{6\ i}r_c^{-6}\right)\right]r_c dr_c,	
\end{multline}
\begin{multline}
\label{eq62}
\widetilde{\delta}=0.356162\mu^{-0.5}T^{-0.5}\int_{r_{c\ min}}^{\infty}\sin{\eta(r_c,\overline{v})} \\  \times \left[1+1.13004 T^{-1}\left(5C_{12\ i}r_c^{-12}-2C_{6\ i}r_c^{-6}\right)\right]r_c dr_c \ .
\end{multline}

\subsection{Exact trajectories}

The intermolecular-distance dependence on time can be developed using the exact solutions of the classical equations of motion. In this case $dt$ and $dr$ are related by \cite{Landau1976}
\begin{equation}
\label{eq63}
dt=\frac{dr}{v\sqrt{1-V^\ast\left(r\right)-\left(r_c/r\right)^2 \left[ 1-V^\ast\left(r_c\right) \right]}}\ .
\end{equation}

\subsubsection{Exact half-trajectories for Lennard-Jones isotropic potentials}
Like the case of parabolic trajectories, the phase shift is given by Eq.~(\ref{eq36}), however the analogues of Eqs.~(\ref{eq47}), (\ref{eq48}) read (in terms of the dimensionless variable $x\equiv r/r_c$)
\begin{multline}
\label{eq64}
\int_{-\infty}^{0}{r_i^{-p}(t)dt}=r_c^{1-p}\overline{v}^{-1} \\ \times \int_{-\infty}^{1}\frac{dx}{x^p\sqrt{1-\frac{8\varepsilon_i\sigma_i^{12}}{\mu\overline{v}^2r_c^{12}x^{12}}+\frac{8\varepsilon_i\sigma_i^6}{\mu\overline{v}^2r_c^6x^6}-\frac{1}{x^2}\left(1-\frac{8\varepsilon_i\sigma_i^{12}}{\mu\overline{v}^2r_c^{12}}+\frac{8\varepsilon_i\sigma_i^6}{\mu\overline{v}^2r_c^6}\right)}}\ \ ,	
\end{multline}
\begin{multline}
\label{eq65}
\int_{0}^{\infty}{r_f^{-p}(t)dt}=r_c^{1-p}\overline{v}^{-1}
\left\{
1-\frac{8\varepsilon_i}{\mu \overline{v}^2}
\left[
\left( \frac{\sigma_i}{r_c}\right)^{12} - \left( \frac{\sigma_i}{r_c}\right)^{6}
\right]
\right\}^{-1/2} \\
\times
\int_{1}^{\infty}\frac{dx}{x^p\sqrt{1-\frac{8\varepsilon_f\sigma_f^{12}}{\mu\overline{v}^2r_c^{12}x^{12}}+\frac{8\varepsilon_f\sigma_f^6}{\mu\overline{v}^2r_c^6x^6}-\frac{1}{x^2}\left(1-\frac{8\varepsilon_f\sigma_f^{12}}{\mu\overline{v}^2r_c^{12}}+\frac{8\varepsilon_f\sigma_f^6}{\mu\overline{v}^2r_c^6}\right)}}
\end{multline}
and lead to
\begin{multline}
\label{eq66}
\eta(r_c)=\frac{1}{\overline{v}}
\left\{
\frac{{\Delta C}_{12}^\prime}{r_c^{11}}
\left[
\int_{-\infty}^{1}\frac{x^{-12}dx}{\sqrt{1-\frac{\widetilde{\xi}_{1i}(r_c)}{x^{12}}+\frac{\widetilde{\xi}_{2i}(r_c)}{x^{6}}-\frac{\left[1-\widetilde{\xi}_{1i}(r_c)+\widetilde{\xi}_{2i}(r_c)\right]}{x^{2}}}} \right. \right. \\
\left.
+\frac{1}{\sqrt{1+{\widetilde{\xi}}_1(r_c)-{\widetilde{\xi}}_2(r_c)}}\int_{1}^{\infty}\frac{x^{-12}dx}{\sqrt{1-\frac{\widetilde{\xi}_{1f}(r_c)}{x^{12}}+\frac{\widetilde{\xi}_{2f}(r_c)}{x^{6}}-\frac{\left[1-{\widetilde{\xi}}_{1f}(r_c)+{\widetilde{\xi}}_{2f}(r_c)\right]}{x^{2}}}}\right]  \\
+\frac{{\Delta C}_6^\prime}{r_c^{11}}\left[\int_{-\infty}^{1}\frac{x^{-6}dx}{\sqrt{1-\frac{\widetilde{\xi}_{1i}(r_c)}{x^{12}}+\frac{\widetilde{\xi}_{2i}(r_c)}{x^{6}}-\frac{\left[1-{\widetilde{\xi}}_{1i}(r_c)+{\widetilde{\xi}}_{2i}(r_c)\right]}{x^{2}}}} \right.
\end{multline}
\[
\left. \left.
 +\frac{1}{\sqrt{1+{\widetilde{\xi}}_1(r_c)-\widetilde{\xi}_2(r_c)}}\int_{1}^{\infty}\frac{x^{-6}dx}{\sqrt{1-\frac{\widetilde{\xi}_{1f}(r_c}{x^{12}}+\frac{\widetilde{\xi}_{2f}(r_c)}{x^{6}}-\frac{\left[1-{\widetilde{\xi}}_{1f}(r_c)+{\widetilde{\xi}}_{2f}(r_c)\right]}{x^{2}}}}\right]\right\} \ ,
\]
where the dimensionless coefficients
\begin{multline}
\label{eq67}
 \widetilde{\xi}_{1i}(r_c)\equiv\frac{2\hbar C_{12i}^\prime}{\mu \overline{v}^2r_{c\ }^{12}}, \ \widetilde{\xi}_{2i}(r_c)\equiv\frac{2\hbar C_{6i}^\prime}{\mu \overline{v}^2r_c^{12}}, \ \widetilde{\xi}_{1f}(r_c)\equiv\frac{2\hbar C_{12f}^\prime}{\mu v^2r_{c\ }^{12}}, \\ \widetilde{\xi}_{2f}(r_c)\equiv\frac{2\hbar C_{6f}^\prime}{\mu v^2r_c^{12}}, \ \widetilde{\xi}_1(r_c)\equiv\frac{2\hbar{\Delta C}_{12}^\prime}{\mu v^2r_{c\ }^{12}}, \ \widetilde{\xi}_2(r_c)\equiv\frac{2\hbar\Delta C_6^\prime}{\mu v^2r_c^{12}}
\end{multline}
are introduced. Re-using the reduced variable $y$ we obtain
\begin{equation}
\label{eq68}
\widetilde{\gamma}=0.3561617 r_{c\ min}^2\mu^{-0.5}T^{-0.5} \int_{1}^{\infty}
\left[
1-\cos{\eta(yr_{c\ min})}
\right] \mathcal{B} (\xi_1,\xi_2,y)dy,
\end{equation}
\begin{equation}
\label{eq69}
\widetilde{\delta}=0.3561617 r_{c\ min}^2\mu^{-0.5}T^{-0.5}\int_{1}^{\infty}{\mathrm{sin}\eta(yr_{c\ min})\mathcal{B}\left(\xi_1,\xi_2,y\right)dy} \ .
\end{equation}

\subsubsection{Exact half-trajectories for numerical potentials}
Taking account of Eqs.~(\ref{eq57}) and (\ref{eq63}), the resulting phase shift is written as
\begin{multline}
\label{eq70}
\eta(r_c)=\frac{1}{\overline{v}}\int_{-\infty}^{r_c}\frac{\Delta V(r)dr}{\sqrt{1-V_i^\ast(r)-(r_c/r)^2(1-V_i^\ast(r_c))}} \\
+\frac{1}{\widetilde{v}}\int_{r_c}^{\infty}\frac{\Delta V(r)dr}{\sqrt{1-V_f^\ast(r)-(r_c/r)^2(1-V_f^\ast(r_c))}}\ ,	
\end{multline}
and further use of Eq.~(\ref{eq41}) leads to
\begin{multline}
\label{eq71}
\widetilde{\gamma}=0.3561617 \mu^{-0.5}T^{-0.5}
\int_{r_{c\ min}}^{\infty}
\left[ 1-\cos{\eta (r_c)} \right] \\
\times \left[ 1-V_i^{\ast}(r_c)-r_cV_i^{\ast \prime}(r_c) /2\right]r_cdr_c \ ,
\end{multline}
\begin{multline}
\label{eq72}
\widetilde{\delta}=0.3561617 \mu^{-0.5}T^{-0.5}
\int_{r_{c\ min}}^{\infty}
\sin{\eta (r_c)} \\
\times \left[ 1-V_i^{\ast}(r_c)-r_cV_i^{\ast \prime}(r_c) /2\right]r_cdr_c
\ .
\end{multline}

\subsection{Applications to NO-Ar and NO-N$_2$}
Since the trajectory is curved by the isotropic potential, this curvature is more pronounced at short intermolecular distances. These distances are probed by collisions of the active molecule with atoms and non-polar perturbers. Moreover, the quality of the interaction potentials was better for NO than for OH. Therefore, we performed a comparative analysis of different models of curved trajectories for two molecular systems NO-Ar and NO--N$_2$. Calculated $\widetilde{\gamma}$ and $\widetilde{\delta}$ values are collected in Tables~\ref{Table5} and \ref{Table6}.

\begin{table*}
\footnotesize
  \caption{\ Comparison of MTVA line-broadening and line-shift coefficients (in cm$^{-1}$atm$^{-1}$) obtained with straight-line, parabolic and exact trajectories for NO-Ar. Lennard-Jones parameters from Fit I are used. Experimental values are from Table~\ref{Table2}}
  \label{Table5}
  \begin{tabular*}{\textwidth}{@{\extracolsep{\fill}}lllllllll}
    \hline
  & & Straight &     & Parabolic & & Exact & & Expt \\
  & & $\Delta V$ LJ & $\Delta V$ num PF1/PF2 & $\Delta V$ LJ & $\Delta V$ num & $\Delta V$ LJ & $\Delta V$ num &  \\
\hline
295~K & $\widetilde{\gamma}$ & 0.206 & 0.568 & 0.209 & 0.224 & 0.207 & 0.182 &  [0.25,0.27] \\
   & $\widetilde{\delta}$ & -0.149 & -0.0005/-0.0003 & +0.142 & -0.081 & -0.147 & -0.040 &  -0.16 \\
    \hline
2800~K & $\widetilde{\gamma}$ & 0.043 & 0.184 & 0.039 & 0.051 & 0.043 & 0.057 &  0.058 \\
   & $\widetilde{\delta}$ & -0.031 & -0.00017/-0.00009 & +0.032 & -0.023 & -0.031 & -0.019 &  -0.043 \\
    \hline
  \end{tabular*}
\end{table*}

\begin{table*}
\footnotesize
  \caption{\ Comparison of MTVA line-broadening and line-shift coefficients (in cm$^{-1}$atm$^{-1}$) obtained with straight-line, parabolic and exact trajectories for NO-N$_2$. Experimental values are from Tables~\ref{Table2} and \ref{Table3}}
  \label{Table6}
  \begin{tabular*}{\textwidth}{@{\extracolsep{\fill}}lllllllll}
    \hline
  & & Straight &     & Parabolic & & Exact & & Expt \\
  & & $\Delta V$ LJ & $\Delta V$ num PF1/PF2 & $\Delta V$ LJ & $\Delta V$ num & $\Delta V$ LJ & $\Delta V$ num &  \\
\hline
295~K & $\widetilde{\gamma}$ & 0.186 & 0.601/0.613 & 0.146 & 0.530 & 0.142 & 0.497 &  [0.28,0.35] \\
   & $\widetilde{\delta}$ & -0.135 & +0.0032/-0.0019 & +0.114 & -0.033 & -0.152 & -0.148 &  [-0.17,-0.18] \\
    \hline
2700~K & $\widetilde{\gamma}$ & 0.039 & 0.198/0.203 & 0.034 & 0.142 & 0.037 & 0.144 &  0.056 \\
   & $\widetilde{\delta}$ & -0.029 & +0.0011/-0.0006 & +0.037 & -0.040 & -0.030 & -0.050 &  -0.052 \\
    \hline
  \end{tabular*}
\end{table*}

A common feature of calculations with traditional straight-line trajectories is that the attempt to improve results by using a numerical treatment of the potential difference $\Delta V$ only worsens the predicted linewidths  (overestimation by 2.5-4.5 times) and shifts (underestimation by several orders of magnitude). In the NO-N$_2$ case the shift sign even becomes positive, in disagreement with the measurements.

The use of curved parabolic trajectories for NO-Ar leads to a substantial improvement in linewidths if the potential difference $\Delta V$ is treated numerically. This numerical $\Delta V$ gives also much more realistic (negative) line shifts (although these are still underestimated by a factor of two), contrary to the Lennard-Jones representations resulting in big and unrealistically positive $\widetilde{\delta}$-values. The positive shifts obtained with Lennard-Jones model potentials can be ascribed to a bad representation of the attractive region, i.e. large intermolecular distances providing the dominant contribution to the line shift. Nevertheless, these ``badly representing'' the attractive region Lennard-Jones parameters appear to be quite realistic if coupled to the exact-trajectory model (which describes more accurately the time dependence of $r$). The use of exact trajectories and the numerical potentials yields worse shifts but an excellent line-width estimate for 2800~K. In the NO-N$_2$ case the use of curved trajectories does not improve theoretical estimates of line-shape parameters. In part, this can be ascribed to a lower quality of the calculated potential energy surfaces.

The attempts to improve the trajectory model within  traditional phase-shift theory show quite fluctuating results. These results confirm the ``internal coherence'' rule: a more accurate than straight-line trajectory description should be accompanied by a higher theory level, as indicated by Szudy and Baylis \cite{Szudy1975}.

\section{Conclusion}

Given the urgent need for pressure-induced line-shape parameters of vibronic transitions for hot-temperature diagnostics and current and future space missions,
we revisited the traditional phase-shift theory with the commonly used model of straight-line trajectories.
This simple classical approach was preferred
because of a much lower, with respect to quantum-mechanical methods, computational cost and sufficiency of order-of-magnitude estimates requested for a huge amount of molecular pairs present (or expected) in hot (exo)planetary atmospheres but inaccessible via laboratory measurements.
We started with a general analysis of pressure-induced linewidths and shifts, using mainly Lennard-Jones 12-6 expressions for the isotropic interaction potentials in the ground and excited electronic states of the active molecule. This analysis was conducted for arbitrary molecular systems, in terms of the dimensionless parameter $\alpha$ determined by the differences of the Lennard-Jones parameters in both states and accounted for various sign sets of the so-called trajectory integrals computed numerically as functions of $\alpha$. We also addressed the temperature dependence of these line-shape parameters and the validity of the commonly used power law, analysing the $\alpha$-dependence of the temperature exponents related to linewidths and shifts, as was shown previously in a similar study by Cybulski et al \cite{Cybulski2013}. The shift-sign change (breaking down of the power law for shifts) observed experimentally for some molecular systems with temperature increasing was also evidenced by varying $\alpha$ from small (leading dispersion interactions and/or low temperatures) to high (leading repulsion and/or high temperatures) values. Moreover, we demonstrated that performing a Maxwell-Boltzmann average on the relative molecular velocities reduces the oscillations of the temperature exponents, i.e. reduces the effects of numerical integration of strongly oscillating functions, in the region around $\alpha \approx 0.1$ and leads to smoother and more physically justified curves.

To check the reliability of the phase-shift theory for representative molecular systems, we chose NO and OH as active molecules (their dipole moments differ by an order of magnitude) and Ar and N$_2$ as perturbers (they give leading dispersive and dipole-quadrupole interactions). These systems were also studied experimentally over a large range of temperatures. To get the Lennard-Jones parameters required in linewidth/shift calculations, we computed the potential energy surfaces in the ground and excited electronic radiator’s states at various collision geometries and extracted the isotropic parts. When the shape of the extracted interaction potential deviated significantly from a Lennard-Jones form  over the range of intermolecular distances considered, we emphasized the repulsive-wall regions (which dominates collisions at high temperatures which are of interest for us) and demonstrated, with an example of NO-Ar, that such a choice leads to much more realistic estimates of line-shape parameters than fits performed on the attraction region
(which is important for low temperatures). Generally, the (absolute) values of line broadening and line-shifting coefficients at room (295~K) and high (2700/2800~K) temperatures were found to be underestimated by not more that about 30\% (except for $\widetilde{\delta}$ of NO-N$_2$ at 2700~K where an underestimate of 44\% was obtained). The quality of predictions was nearly the same for perturbation of NO by Ar and nitrogen. For OH, there were not enough measurements to draw a definite conclusion in the OH-Ar case, and, moreover, difficulties with the potential energy calculations for OH-N$_2$ prevented us from computing linewidth and shift estimates. On the basis on the molecular systems considered, we can conclude that the traditional phase-shift theory provides rather underestimated absolute values of broadening and shift coefficients with uncertainties better than 50\%. This means that for cases were just an order of magnitude is required, the traditional theory can be used.

To avoid the fragility of this approach related to the high sensitivity of the Lennard-Jones parameters to the choice of the intermolecular-distance region selected for fits, we also tested a numerical representation of the potential difference $\Delta V$ for NO-Ar, completed by tests of extrapolation to the region of very small $r$-values. Finally, the extrapolation type was found to be not important for linewidths (because of strong oscillations of the cosine function giving nearly zero contribution from this region), but the broadening coefficient values at both temperatures  considered were much bigger than the measurements. For line shifting, although the correct negative signs were reproduced, underestimations of absolute values by several orders of magnitude were observed. Therefore, an ``isolated'' improvement of the potential in the frame of the traditional phase-shift theory with rectilinear trajectories was inefficient to get more realistic line-shape estimates.

As a next step, we tried (simultaneous with and separate from the potential description) improvements of the trajectory model, using for simplicity the mean-thermal-velocity approximation. Two types of trajectories curved by the force derived from the isotropic potential were considered: parabolic and exact ones, assuming that the two halves of each trajectory are driven, respectively, by the isotropic potentials in the initial and final radiator’s states. Applications made for the NO-Ar and NO-N$_2$ systems demonstrated quite disparate predictions, so we can also conclude that the trajectory improvements are still insufficient to match completely the measurements.

The results of our two improvement attempts (numerical potentials and curved trajectories) support the conclusion made by Szudy and Baylis \cite{Szudy1975} that going beyond the straight-line trajectories requires at least a first-order correction to the usual phase-shift integral and significant differences with respect to the standard approach can be expected. Developing such a formalism represents a serious independent work and will be addressed in a future study. Another subject of future investigations could be the rotational dependence of the line-shape parameters, which is generally neglected by the theory but still detectable experimentally for some molecular systems; considering this dependence will require an account of anisotropic interactions.


\section*{Conflicts of interest}
There are no conflicts to declare.

\section*{Acknowledgements}

This work was supported by the European Research Council under Advanced Investigator Project 883830 and by UK STFC under grant ST/R000476/1. The authors acknowledge the use of the UCL Myriad High Performance Computing Facility and associated support services in the completion of this work. JB acknowledges the mobility grant provided by the Region Bourgogne-Franche-Comt\'{e} for her visit to UCL.



\appendix
\section{Appendix} \label{appendix_a}

\setcounter{table}{0}
\renewcommand{\thetable}{A\arabic{table}}
\renewcommand*{\theHtable}{\thetable}

\begin{table*}[!ht]
\tiny
    \centering
    \caption{Definitions of parameters, integration variables and trajectory integrals for 12-$m$ Lennard-Jones interactions ($m$~=~6, 5, 4). Both initial and practically used for numerical integration expressions are given for $B(\alpha)$ and $S(\alpha)$}
    \label{TableA1}
    \begin{tabular}{lcccc}
            \hline
            Potential Form & $\alpha$ & $x$ & $B(\alpha)$ & $S(\alpha)$ \\
\hline
12--6 &
$\frac{7}{2^{7/5}3^{1/5}\pi^{6/5}}{\overline{v}}^{6/5}\frac{\Delta C_{12}^{\prime}}{\left| \Delta C_{6}^{\prime} \right|^{11/5}} $ &
$b\left( \frac{8\overline{v}}{3\pi\left| \Delta C_{6}^{'} \right|} \right)^{1/5}$ &
$\int_{0}^{\infty}\text{sin}^{2} \left(\frac{1}{2}(\alpha x^{- 11} - x^{- 5})\right)xdx$ & $\int_{0}^{\infty}\sin(\alpha x^{- 11} - x^{- 5})xdx$ \\
 & $\approx 0.5391\ {\overline{v}}^{6/5} \Delta C_{12}^{\prime}\left| \Delta C_{6}^{\prime} \right|^{- 11/5}$ &  &
$\frac{1}{4}\int_{0}^{\infty}\sin (\alpha x^{- 11} - x^{- 5} )(11\alpha x^{- 10} - 5x^{- 4}) dx $ &
$\frac{1}{2}\int_{0}^{\infty}\cos( \alpha x^{- 11} - x^{- 5})(11\alpha x^{- 10} - 5x^{- 4}) dx$ \\
\hline
12-5 &
$\frac{7\pi 3^{19/4}}{2^{27/2}}{\overline{v}}^{7/4}\frac{\Delta C_{12}^{\prime}}{\left| \Delta C_{5}^{\prime} \right|^{11/4}}$ &
$b\left( \frac{3\overline{v}}{4\left| \Delta C_{5}^{\prime} \right|} \right)^{1/4}$ &
$\int_{0}^{\infty}\sin^{2}\left(\frac{1}{2}(\alpha x^{- 11} - x^{- 4})\right)xdx$ &
$\int_{0}^{\infty}\sin(\alpha x^{- 11} - x^{- 4})xdx $ \\
& $\approx 0.3505\ {\overline{v}}^{7/4} \Delta C_{12}^{\prime}\left| \Delta C_{5}^{\prime} \right|^{- 11/4}$ & &
$\frac{1}{4}\int_{0}^{\infty}\sin( \alpha x^{- 11} - x^{- 4})(11\alpha x^{- 10} - 4x^{- 3}) dx$ &
$\frac{1}{2}\int_{0}^{\infty}\cos(\alpha x^{- 11} - x^{- 4})(11\alpha x^{- 10} - 4x^{- 3}) dx$ \\
\hline
12-4 &
$\frac{7 \cdot 3^{2}}{2^{13/3}\pi^{8/3}}{\overline{v}}^{8/3}\frac{\Delta C_{12}^{\prime}}{\left| \Delta C_{4}^{\prime} \right|^{11/3}}$ &
$b\left( \frac{2\overline{v}}{\pi\left| \Delta C_{4}^{\prime} \right|} \right)^{1/3}$ &
$\int_{0}^{\infty} \sin^{2} \left(\frac{1}{2}(\alpha x^{- 11} - x^{- 3})\right)xdx$ &
$\int_{0}^{\infty}\sin(\alpha x^{- 11} - x^{- 3})xdx$ \\
& $\approx 0.1476\ {\overline{v}}^{8/3} \Delta C_{12}^{\prime}\left| \Delta C_{4}^{\prime} \right|^{- 11/3}$ & &
$\frac{1}{4}\int_{0}^{\infty}\sin ( \alpha x^{- 11} - x^{- 3} )(11\alpha x^{- 10} - 3x^{- 2}) dx$ &
$\frac{1}{2}\int_{0}^{\infty}\cos(\alpha x^{- 11} - x^{- 3})(11\alpha x^{- 10} - 3x^{- 2}) dx$ \\
            \hline
        \end{tabular}
\end{table*}

\begin{table}[!ht]
\footnotesize
    \centering
    \caption{Linewidth and shift expressions for 12-\emph{m} Lennard-Jones interactions (\emph{m}~=~6, 5, 4); see Table A1 for the corresponding expressions of trajectory integrals.}
    \label{TableA2}
    \begin{tabular}{lcc}
            \hline
            Potential Form & $\gamma/N$ & $\delta/N$ \\
            \hline
12--6 &
$2\left( \frac{3\pi}{8} \right)^{2/5}\left| \Delta C_{6}^{\prime} \right|^{2/5}{\overline{\text{v}}}^{3/5}B(\alpha)$
&
$\left( \frac{3\pi}{8} \right)^{2/5}\left| \Delta C_{6}^{\prime} \right|^{2/5}{\overline{\text{v}}}^{3/5}S(\alpha)$ \\
12-5 &
$2\left( \frac{4}{3} \right)^{1/2}\left| \Delta C_{5}^{\prime} \right|^{1/2}{\overline{\text{v}}}^{1/2}B(\alpha)$
&
$\left( \frac{4}{3} \right)^{1/2}\left| \Delta C_{5}^{\prime} \right|^{1/2}{\overline{\text{v}}}^{1/2}S(\alpha)$ \\
12-4 &
$2\left( \frac{\pi}{2} \right)^{2/3}\left| \Delta C_{4}^{\prime} \right|^{2/3}{\overline{\text{v}}}^{1/3}B(\alpha)$
&
$\left( \frac{\pi}{2} \right)^{2/3}\left| \Delta C_{4}^{\prime} \right|^{2/3}{\overline{\text{v}}}^{1/3}S(\alpha)$ \\
            \hline
        \end{tabular}
\end{table}






\end{document}